\pdfoutput=1
\documentclass[aip,jcp,preprint,amsmath,amssymb,showpacs,mathptmax]{revtex4-1}
\usepackage{graphicx,color,hyperref}

\usepackage[version=1]{mhchem}
\usepackage{mathptmx}
\usepackage{times}
\usepackage{xcolor}
\usepackage{epsfig,amsopn}
\usepackage{graphicx}
\usepackage{bm,amsmath,amssymb}
\usepackage{natbib}
\usepackage{braket}
\usepackage{float}
\usepackage{enumerate}
\usepackage{hyperref}
\usepackage[export]{adjustbox}
\allowdisplaybreaks[1]

\usepackage{color}
\definecolor{dgreen}{rgb}{0,0.7,0}

\begin{document}
\def\be{\begin{equation}}
\def\ee{\end{equation}}
\def\bea{\begin{eqnarray}}
\def\eea{\end{eqnarray}}
\def\f{\frac}
\def\l{\label}
\def\nn{\nonumber}

\newcommand{\eref}[1]{Eq.~(\ref{#1})}%
\newcommand{\Eref}[1]{Equation~(\ref{#1})}%
\newcommand{\fref}[1]{Fig.~\ref{#1}} %
\newcommand{\Fref}[1]{Figure~\ref{#1}}%
\newcommand{\sref}[1]{Sec.~\ref{#1}}%
\newcommand{\Sref}[1]{Section~\ref{#1}}%
\newcommand{\aref}[1]{Appendix~\ref{#1}}%
\newcommand{\sgn}[1]{\mathrm{sgn}({#1})}%
\newcommand{\erfc}{\mathrm{erfc}}%
\newcommand{\erf}{\mathrm{erf}}%

\title{Motion of a Brownian molecule in the presence of reactive boundaries}
\author{Arnab Pal}
\affiliation{School of Chemistry, Raymond and Beverly Sackler Faculty of Exact Sciences, Tel Aviv University, Tel Aviv 69978, Israel}
\affiliation{Center for the Physics and Chemistry of Living Systems. Tel Aviv University, 6997801, Tel Aviv, Israel}
\affiliation{The Sackler Center for Computational Molecular and Materials Science, Tel Aviv University, 6997801, Tel Aviv, Israel}
\author{Isaac P\'erez Castillo}
\affiliation{Department of Quantum Physics and Photonics, Institute of Physics, UNAM, P.O. Box 20-364, 01000 Mexico City, Mexico}
\affiliation{London Mathematical Laboratory, 14 Buckingham Street, London WC2N 6D, United Kingdom}
\author{Anupam Kundu}
\affiliation{International Centre for Theoretical Sciences, TIFR, Bangalore 560089, India}

\date{\today}

\begin{abstract}
We study the one-dimensional motion of a Brownian particle inside a confinement 
described by two reactive boundaries which can  partially reflect
or absorb the particle. Understanding the effects of such boundaries is 
important in physics, chemistry and biology. 
We compute the probability density of the particle displacement
exactly, from which we derive expressions for the survival probability and the
mean absorption time as a function of the reactive coefficients. Furthermore, using
the Feynman-Kac formalism, we investigate the reaction time profile, which is the
fluctuating time spent by the particle at a given location, both till a fixed observation 
time and till the absorption time. Our analytical results are compared to numerical simulations showing perfect agreement.
\end{abstract}

\maketitle


\section{Introduction}
\label{sec:Introduction}

Diffusion is a paradigm of stochastic processes that successfully provides a basic description of 
various phenomena like chemical reactions, or bio-molecular processes occurring  at
cellular and sub-cellular levels \cite{Berg83,Crank75}. For instance in molecular
biology, the motion of a protein molecule in the solution inside a 
living cell can be considered as a simple diffusion \cite{Berg83, Crank75, Cooper}.
The protein molecules perform  everlasting motions
due to their thermal energy and, as a result,   the trajectory of the protein molecule is erratic
and the density of the molecules slowly spreads throughout the medium.
While diffusion can predict motion of protein molecules
well inside the cellular domain
under dilute conditions, the behavior
gets affected by the nature of the boundaries \cite{Berg83, Crank75, Cooper}. 
This could be due to the
structure of the cell membrane that protects and organizes cells by 
regulating not only what enters or exits
the cell, but also by how much \cite{Vroman,Bychuk}. 
In this paper we examine such gate keeping functionalities of the boundaries (semi-permeable or resistive in nature)
which optimally control the flow of essential chemical species across  the cellular  membrane \cite{Grebenkov06,GrebenkovRMP07}.

\begin{figure}[t]
\includegraphics[scale=0.48]{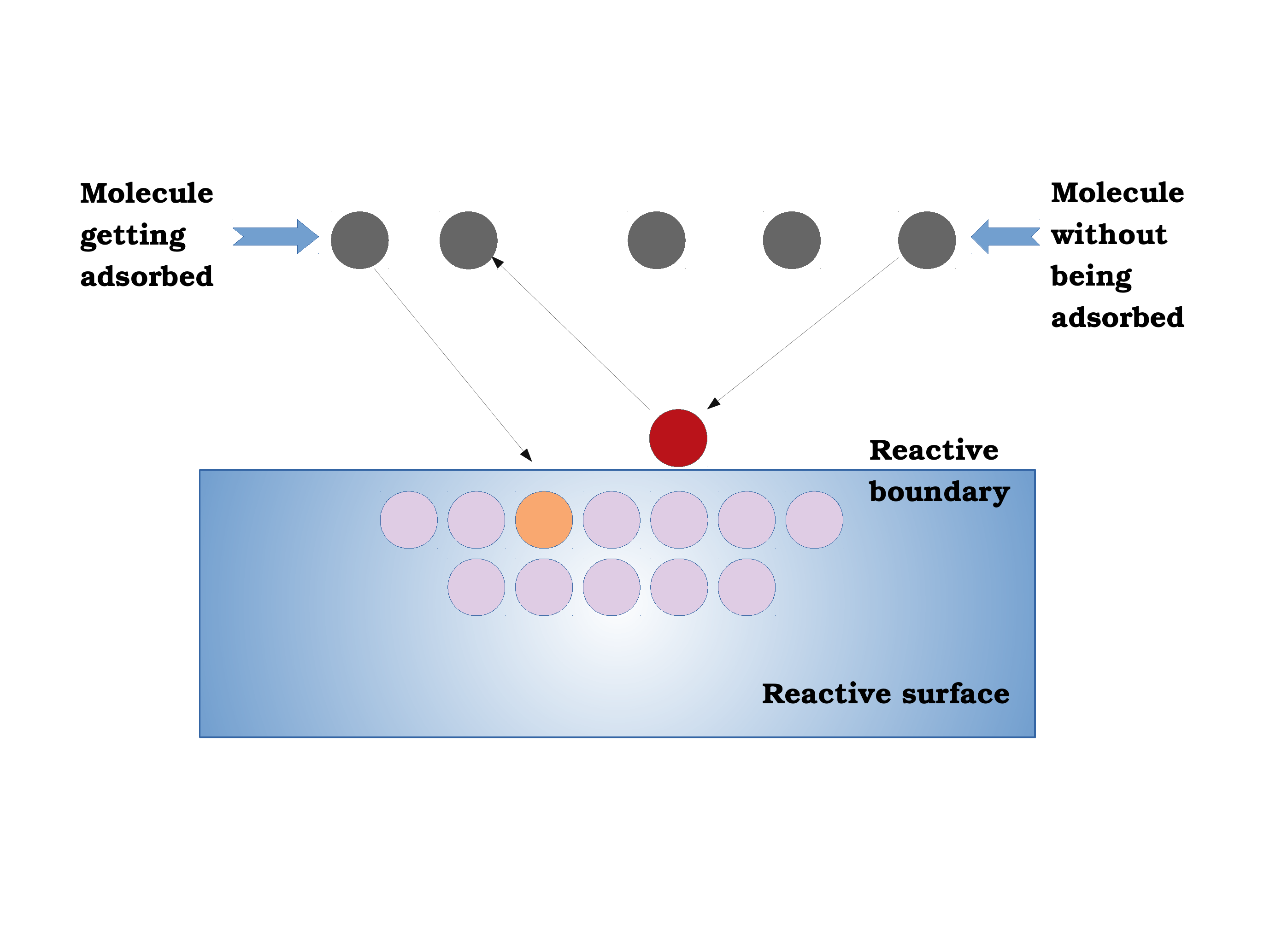}
\caption{Schematics of a chemically reactive surface. Some Brownian molecules are  being adsorbed by the constituents of  a reactive surface.
Others react with the constituents at the reactive boundary, forming volatile compounds (indicated in red) before being released to the environment.
Reactive constants will characterize the nature of such surfaces. This adsorbing phenomena corresponds to a partially absorbing or reactive surface.}
\label{fig-cart}
\end{figure}

The simplest  types of boundary conditions can
be formulated in terms of  either vanishing flux through the boundary (usually called reflecting or impermeable boundary)
or vanishing density   at the
boundary (called absorbing  boundary) \cite{Zwanzig90,Allison85,Lamm83,Chapman16,Vrentas89,Erban07,
Hattne05,Naqvi82,Jacob-2012,El-Sh-2000,Goodrich-54,Grebenkov-review,Isaac}.
In the first case, a diffusing molecule is reflected
whenever it hits the boundary, while in the second type of boundary condition a diffusing molecule is removed from the system
whenever it hits the boundary, which can be interpreted  as the molecule being absorbed at the boundary.
However, more
realistic boundary conditions can be realized in terms of a partially absorbing
boundary (also termed as Robin, radiative or mixed boundary conditions\cite{Crank75}), 
which means that  a molecule may be absorbed (or reflected) with some probability
\cite{Hippel,Slutsky,Grebenkov15,Dickman01,Skorokhod59,Burdzy04,Singer08,Batsilas03,Berezhkovskii,Monine} (see  \fref{fig-cart}).
 From a 
bio-chemical point of view, this absorption probability depends on the  reactivity of the boundary (e.g. on the rate constant of the
adsorbing chemical reaction and on the number of available receptors), and on the details of the
model. The reactivity constant can also be measured experimentally from the chemical
properties of the boundary (see e.g. \cite{GrebenkovRMP07} and references therein).

It is worth mentioning that the interaction of a diffusive particle with a reactive
boundary is also of practical importance, 
since they offer plentiful industrial applications
in surface or colloid science,  and materials research \cite{Metzler07}. 
 Few examples worth mentioning are:  fluid or mass transport in porous media \cite{Stapf06},
electric transport in electrolytic cells \cite{Sapoval5},
nuclear magnetic resonance (diffusion of spins in
confining porous media), and
applications  to foam relaxation and
surfactants\cite{Sonin}. Other examples can be found in physiology where
oxygen molecules can penetrate across alveolar membranes
for further adsorption in blood, or are bounced back and continue
the motion. The proportion of adsorbed and reflected oxygen molecules can be 
characterized by permeability varying from zero (perfectly reflecting boundary) to 
infinity (perfectly adsorbing boundary) \cite{Weibel,Filoche,Sapoval1}. 
A similar description can be useful to explain heterogeneous
catalysis frequently observed in petrochemistry e.g. chemical vapor decomposition or plasma etching \cite{Vrentas89}.
The reactive molecules are injected 
into a solvent and then they diffuse towards a catalyst. Hitting the catalytic
surface, they can be either transformed into other molecules (with a finite reaction
rate), or  are bounced back for further diffusion in the bulk \cite{Sapoval2,Sapoval3,Sapoval4}.

In this paper we investigate the motion of non-interacting diffusing molecules
inside a reactive domain. 
If left alone the molecules may eventually decay or get adsorbed
at the boundary of the confining domain after some time. This is called the lifetime 
when the molecule gets adsorbed. Clearly, this time is a random quantity whose cumulative probability, 
called the survival probability, simply measures the chance for the species to remain inside a confining
domain up to a fixed time $t$ without being adsorbed or decay. 
In  the literature, computing  this distribution is 
known as the first passage time problem and  it has been the subject of interest to scientists 
for many decades \cite{BMS13,Majumdar:2005,Redner,First-Passage-Book,Benichou2011RMP}. 
First passage time problems have ubiquitous applications
in physical, biological and chemical processes, ranging from  finance to animal foraging theory.
Few examples are: survival time of a bacteria to remain alive while searching for food, 
average lifetime of a messenger RNA which is translated into protein by the joint action of transfer RNA (tRNA) and the ribosome,
binding time of a protein to an enzyme, search time of animals for food resources, etc. 
\cite{BMS13,Majumdar:2005,Redner,First-Passage-Book,Benichou2011RMP}.
Motivated by this backdrop, we investigate the survival properties of diffusive particles inside a 
reactive domain \cite{Grebenkov17, Salminen11}.
In addition, we are also interested in the extreme displacements made by the diffusing particle and time spent per unit
length around a spatial point in the presence of reactive boundaries. This time density is  also called the local  or reaction time. 
These quantities are important in characterizing the motion of the molecule. Indeed, the extreme displacement describes
the geometrical properties of the trajectories of the molecule, while the  reaction time describes the temporal
distribution of the trajectories over space  and, as such, captures the time spent by a molecule nearby a reactive agent placed at a
specific region of the space, upon which the reaction  takes place.

We here provide a brief summary of our results which can be divided into three parts. In the first part,
we  obtain exact analytical results for the propagator, survival probability, and the mean adsorption time. 
In the second part, we compute the distribution of the maximum displacement of the molecule by the method of counting paths.
In the third part of the paper we  study the statistical properties of the reaction time   using the Feynman-Kac method 
of Brownian functionals in two cases: (i) when the observation
time is fixed, and (ii) when the observation time is random.  In all cases, we have verified our 
results  by contrasting them to numerical simulations.

The paper is organized as follows. In \sref{Definitions}, we  introduce the 
working model and all the relevant observables which we have studied in subsequent sections.
In \sref{Results}, we provide the details of the derivations and present the results. 
In particular, general expressions for the propagator, survival probability and the mean adsorption time  are given
in \sref{Results-propagator}. These results are used to
compute the full statistics of the maximum displacement in \sref{max-displacement}. In \sref{reaction-time-profile}
we investigate the reaction time profile and present associated exact results.
Some of the detailed derivations have been relegated to the appendix \sref{appendix} for clarity. We conclude
our paper in \sref{conclusion}.

\section{Definitions and notations of the quantities of interest}
\l{Definitions}
\noindent
 Here we provide the definitions and  notations of the quantities that we are interested to study  and while doing so we also briefly review the basic concepts and interpretations  in the context of a Brownian molecule in presence of reactive boundaries. 
Imagine that we want to understand the motion of a protein molecule, e.g. an enzyme inside a cell, whose motion can be well described  as a diffusion process, in which 
the boundaries of the cell can be approximated as reactive boundaries.  For simplicity we consider the one-dimensional case in which the particle diffuses inside a domain $x\in[0,L]$.
Mathematically, this process is described by a propagator $G_L(x,x_0,t)$ which simply represents 
the probability density for  a molecule to be found at $x$ at time $t$ given that it  started at $x_0$ at an initial time $t_0=0$. It can be shown that the propagator satisfies the diffusion equation subject to the reactive boundary conditions at $x=0$ and $x=L$:
\bea
\frac{\partial}{\partial t}G_L(x,x_0,t)&=&D\frac{\partial^2}{\partial x^2}G_L(x,x_0,t) \l{propagator-reactive-BC}\,,  \\
\frac{\partial}{\partial x}G_L(0,x_0,t)&=&\alpha_0 G_L(0,x_0,t) \l{propagator-reactive-BC-I}\,, \\
\frac{\partial}{\partial x}G_L(L,x_0,t)&=&\alpha_L G_L(L,x_0,t) \l{propagator-reactive-BC-II}\,,
\eea
 where $D$ is the diffusion constant. For clarity, we will consider deterministic initial conditions, that is, $G_L(x,x_0,0)=\delta(x-x_0)$. 
 The parameter $\alpha_0$ (resp. $\alpha_L$) controls how often a molecule hitting the boundary at $x=0$ (resp. $x=L$) will be either reflected or absorbed. By tuning these values one can go from a perfectly reflecting boundary to a perfectly absorbing boundary.  
The problem of finding the propagator have been considered earlier in several contexts
mostly in a semi infinite space with one reactive boundary, for example, with step initial condition for the concentration \cite{Ben-naim93}, in target search 
 problems \cite{Whitehouse-Martin-Satya}, or in diffusion controlled recombinations \cite{Sano79}.
It has been shown that solving the diffusion equation in a bounded domain with reactive boundary conditions is equivalent to 
solving the diffusion equation in unbounded domain with ``sink" terms \cite{Ben-naim93, Redner}.

From the propagator $G_L(x,x_0,t)$, the survival probability 
that  a molecule has not been absorbed or decomposed till the observation time $t$  
is simply given by
\bea
S_L(x_0,t)=\int_{0}^{L}~dx~G_L(x,x_0,t)~.
\l{Full-survival}
\eea
A simple interpretation of this expression comes from a path counting  argument and works as follows. 
The propagator  contains the contributions  from all  
 the statistical paths that start at time $t_0=0$ at position a $x_0$ and end at time $t$ at position a $x$ without being absorbed at
 either boundary. There are four types of such paths: those which have never  reached either 
boundary at $x=0$ and $x=L$ in time $t$, those which may have hit one of the boundaries but got reflected 
and those which may have hit both boundaries and, again,  got reflected. The survival
probability then gets contribution from all such paths which reach any final point $x\in (0,L)$.
 Analogously to the concept of the first
passage time, one can introduce the first absorption time $t_a$, being the time at which a 
molecule is absorbed at either boundary. If we denote its probability density as $f_L(x_0,t)$, it is easy to see that $S_L(x_0,t)=\text{Prob.}(t_a>t|x_0)=\int_0^tdt'f_L(x_0,t')$, which implies $f_L(x_0,t)=-\frac{dS_L(x_0,t)}{dt}$.

A related important quantify that we are interested in is the mean absorption time $T_L(x_0)$
(MAT), which can be computed from the density $f_L(x_0,t)$ as 
\bea
T_L(x_0)=\int_{0}^{\infty}~dt~t f_L(x_0,t).
\l{MFPT}
\eea
The MAT  is often considered to be a hallmark quantity due to its ubiquity in problems as diverse as the
average search time for a bacterium to find its food, or the mean turnover time to complete a reaction, or the   running 
time of a computer programming. We refer to \cite{BMS13,Majumdar:2005,Redner,First-Passage-Book,Benichou2011RMP}
for  a comprehensive review  on this subject.

Inside a cellular domain, protein molecules will  perform short excursions
before they either react to substrates or are absorbed at the boundaries. 
The nature of  these paths depend on several factors such as cell concentration, local density of the surrounding
molecules, etc. In this biological scenario, very long trajectories  may be detrimental for
chemical reactions to occur but can be useful when a longer lifetime of
a molecule (i.e. with small absorption at the boundaries) is favored.  A qualitative geometric characterization of  these trajectories
can be provided by the statistics of the molecule's longest excursion, which is, in fact, the maximum displacement of the Brownian particle till a fixed time~$t$. 

With this in mind, let us then consider a simpler situation where a molecule is moving on the positive axis
with a reactive boundary at the origin $x=0$. Let $M$ be the maximum displacement made by the molecule. 
The cumulative probability that $M$ is less than $L$ is denoted by 
\bea
&&H(L,t|x_0)=\text{Prob}[M \leq L,t|x_0] \nonumber \\ 
&&=\text{Prob.}\left[\{x(t') < L;~0\leq t' \leq t\}~|~\text{the~molecule~did~not~get~absorbed~at~the~reactive~boundary}~x=0\right] \nonumber \\ 
&&=\frac{\text{Prob.}[\text{In~time~{\it t}~the~molecule~did~never~hit~}x=L~\text{and~did~not~get~absorbed~at~}~x=0]}{\text{Prob.}[\text{In~time~{\it t}~the~molecule~did~not~get~absorbed~at~}x=0]} \label{mc-2} \\
&&= \frac{\mathcal{S}_L(x_0,t)}{\mathcal{S}_{\infty}(x_0,t)},
\l{maximum-cumulative}
\eea
 where $\mathcal{S}_L(x_0,t)=\int_0^L~\mathcal{G}_L(x,x_0,t)$ is the survival probability of the molecule in the presence of a 
full absorbing boundary at $L$ in addition to a reactive boundary at $x=0$, while $\mathcal{G}_L(x,x_0,t)$ is the propagator which describes such a system. Therefore,
$\mathcal{G}_L(x,x_0,t)=\underset{\alpha_L \to \infty}{\lim} G_L(x,x_0,t) $ and  where we have denoted
$\mathcal{S}_{\infty}(x_0,t)=\underset{L \to \infty}{\lim} 
\mathcal{S}_L(x_0,t)$. 
The above definition comes from a very simple path counting argument.
The cumulative probability $H(L,t|x_0)$ gets contribution from all the paths which start from $x_0$ and reach 
somewhere within $x\in(0,L)$ (while staying below $x=L$ throughout) along with the condition that they survived the reactive boundary at $x=0$
till time $t$. Hence this probability is exactly the fraction of paths of duration $t$ that
starting from $x_0$ never hit $x=L$ among those paths which survive till time $t$ from the reactive boundary at $x=0$. 
In  Sec. \ref{Results-propagator} we compute the cumulative probability of the maximum distance traveled by the molecule.

Having reached the desired active site,
the protein molecule (e.g. enzyme) reacts with substituents or ligands. For example
in the kinetics of an enzymatic reaction mechanism, an enzyme binds to a substrate  
to form a complex, which  in turn releases a product, regenerating the original enzyme.
This kind of reaction scheme 
is due to the pioneering work of Michaelis and Menten who 
further explained how reaction rates depend on the concentration of the enzyme and the substrate \cite{Michaelis-Menten}.
The reaction or binding time of such process is very relevant in biochemistry since prior knowledge could help improve the
efficiency of a chemical reaction through catalysis or by facilitating metabolic pathways. 
A quantitative definition of this time can be formulated as the following
\bea
L_t(y_0,x_0)=\int_0^t~dt'~\delta[ x(t')-y_0 |x_0]~,
\l{local-time}
\eea
which measures the amount of time an enzyme spends around the substrate 
(located at a given coordinate $y_0$) over an interval $0 \leq t'\leq t$. 
By construction, this is a functional of the trajectory and normalized as $\int dy_0~L_t(y_0,x_0)=t$.
 In the theory of stochastic processes, this is often termed
as the local time in diffusion processes \cite{Majumdar:2005,Sanjib-Satya} or the 
empirical density (when appropriately rescaled by the observation time) in generic Markov processes \cite{Donsker-Varadhan,Barato}. 
 In section \sref{reaction-time-profile} we will use the Feynman-Kac path integral formalism to investigate the statistical properties of this Brownian functional.

\section{Main Results}
\l{Results}
\subsection{Propagator, survival probability and the mean absorption time}
\l{Results-propagator}
This section contains our main results on the quantities discussed in the previous section and their explicit derivations. 
To compute these quantities we first need to solve  \eref{propagator-reactive-BC} to find the propagator inside
the domain  $x\in [0,L]$ satisfying the reactive boundary conditions \eref{propagator-reactive-BC-I} and
\eref{propagator-reactive-BC-II} and the initial condition 
\begin{equation}
G_L(x,x_0,t=0)=\delta(x-x_0). \l{G-ini}
\end{equation} 
Applying the method of separation of variables in \eref{propagator-reactive-BC} we can write the propagator in the following way
\bea
G_L(x,x_0,t)=\underset{k}{\sum}~c_k~\psi_k(x)~\psi_k(x_0)~e^{-Dk^2~t},\l{propagator-reactive-BC-initial}
\eea
where  the $k$'s are the eigenvalues whose corresponding eigenfunctions $\psi_k(x)$ satisfy 
\bea
\frac{\partial^2 \psi_k(x)}{\partial x^2} = -k^2 \psi_k(x), \l{psi_k-eq}
\eea
with the boundary conditions (BCs) 
\bea
\frac{\partial}{\partial x}\psi_k(x,t)\Big|_{x\to 0}&=&\alpha_0 \psi_k(x,t)\Big|_{x\to 0}\,, \l{psi-BC-I} \\
\frac{\partial}{\partial x}\psi_k(x,t)\Big|_{x\to L}&=&\alpha_L \psi_k(x,t)\Big|_{x\to L}\,. \l{psi-BC-II}
\eea
A general solution of \eref{psi_k-eq} is given by 
\bea
\psi_k(x)=a(k) \cos(kx)+b(k) \sin(kx). \l{psi-g-s}
\eea
To determine the functions $a(k)$ and $b(k)$ we insert $\psi_k(x)$ back into Eqs. \eqref{psi-BC-I} and \eqref{psi-BC-II} to  obtain
\bea
kb(k)&=&\alpha_0a(k), \l{akbk-1} \\
\tan(kL)&=& \frac{kb(k)-\alpha_La(k)}{ka(k)+\alpha_Lb(k)} \l{akbk-2}\,.
\eea 
Now, assuming  that $a(k)=kf(k)$ and $b(k)=\alpha_0 f(k)$
for some $f(k)$ to be determined from normalization, we have from \eref{akbk-2}  that
\bea
e^{2ikL}=\f{(k+i\alpha_0)(k-i\alpha_L)}{(k-i\alpha_0)(k+i\alpha_L)},
\l{k-equation}
\eea
and from \eref{psi_k-eq}
\bea
\psi_k(x)=f(k)~[k \cos(kx)+\alpha_0 \sin(kx)]. \l{psi-g-s-1}
\eea
Hence, the full solution reads
\bea
G_L(x,x_0,t)=\underset{k\in\mathcal{R}_k(L) }{\sum}~C_k~[k \cos(kx)+\alpha_0 \sin(kx)]~[k \cos(kx_0)+\alpha_0 \sin(kx_0)]~e^{-Dk^2~t},\l{propagator-reactive-BC-initial-1}
\eea
where $C_k=c_k f^2(k)$ is a constant to be determined,
and $\mathcal{R}_k(L)$  is the set of eigenvalues, solution of the transcendental equation
\eqref{k-equation}, for a fixed value of $L$. In the context of finding  the reunion probability of $N$ Brownian particles 
moving on a line with partially absorbing/reflecting crossing conditions, a general multi-particle propagator  has been computed using Bethe ansatz\cite{Isaac}.
The constant $C_k$ in \eref{propagator-reactive-BC-initial-1} can now be found using the initial condition 
in  \eref{G-ini}
\bea
C_k \psi_k(x_0)&=& \f{\int_0^L~dx~ G_L(x,x_0,0)\psi_k(x) }{\int_0^L~dx~ \psi_k(x)\psi_k(x)}\,,
\eea
where we have assumed that the  set of normalized eigenfunctions $\psi_k(x)$ forms 
a complete basis in $[0,L]$ with $k \in \mathcal{R}_k(L)$.  Performing a lengthy manipulation we  obtain the normalization to be 
\bea
C_k=(\alpha_0^2+k^2)^{-1}\left[L+\f{\alpha_0}{\alpha_0^2+k^2}-\f{\alpha_L}{\alpha_L^2+k^2}\right]^{-1}~.
\eea
Substituting this  result in \eref{propagator-reactive-BC-initial-1}, we arrive at the final expression for the full propagator 
\bea
G_L(x,x_0,t)&=&\underset{k\in\mathcal{R}_k(L)}{\sum}~e^{-Dk^2t}~\frac{[k \cos(kx)+\alpha_0 \sin(kx)]~[k \cos(kx_0)+\alpha_0 \sin(kx_0)]}{(\alpha_0^2+k^2)\left[L+\f{\alpha_0}{\alpha_0^2+k^2}-\f{\alpha_L}{\alpha_L^2+k^2}\right]}\,.
\l{propagator-reactive-BC-full}
\eea
Various limits can be immediately examined from \eref{propagator-reactive-BC-full}. For example, in the case
of a semi-infinite domain (where the boundary at $x=L$ is taken to infinity)
the propagator takes the following form (see \sref{propagator in semi-infinite domain} for details)
\bea
G^{\text{si}}(x,x_0,t)=\f{1}{4\alpha_0}~e^{D \alpha_0^2 t}~\mathcal{D}_x\mathcal{D}_{x_0} ~
\Big[ \phi(x-x_0,t)-\phi(x+x_0,t) \Big]~,
\l{propagator-reactive-BC-semi-infinite}
\eea
where  $\mathcal{D}_y= \Big(\f{\partial}{\partial y}+\alpha_0\Big)$, and the function $\phi$ is defined as follows
\bea
\phi(z,t)=e^{\alpha_0 z} \erfc \Big[\f{2D \alpha_0 t+z}{\sqrt{4Dt}}\Big]~ + ~e^{-\alpha_0 z} \erfc \Big[\f{2D \alpha_0 t-z}{\sqrt{4Dt}} \Big].
\l{psi-definition}
\eea
In \fref{propagators}, we have compared our analytical results to Monte Carlo simulations, which were performed according to the method explained in  
\sref{simulations}. The left panel, corresponds to taking a finite size interval, while the right panel is the result of considering a semi-infinite domain. In both 
cases the agreement between theory and Monte Carlo simulations is excellent.

\begin{figure}[t]
\includegraphics[width=.4\hsize]{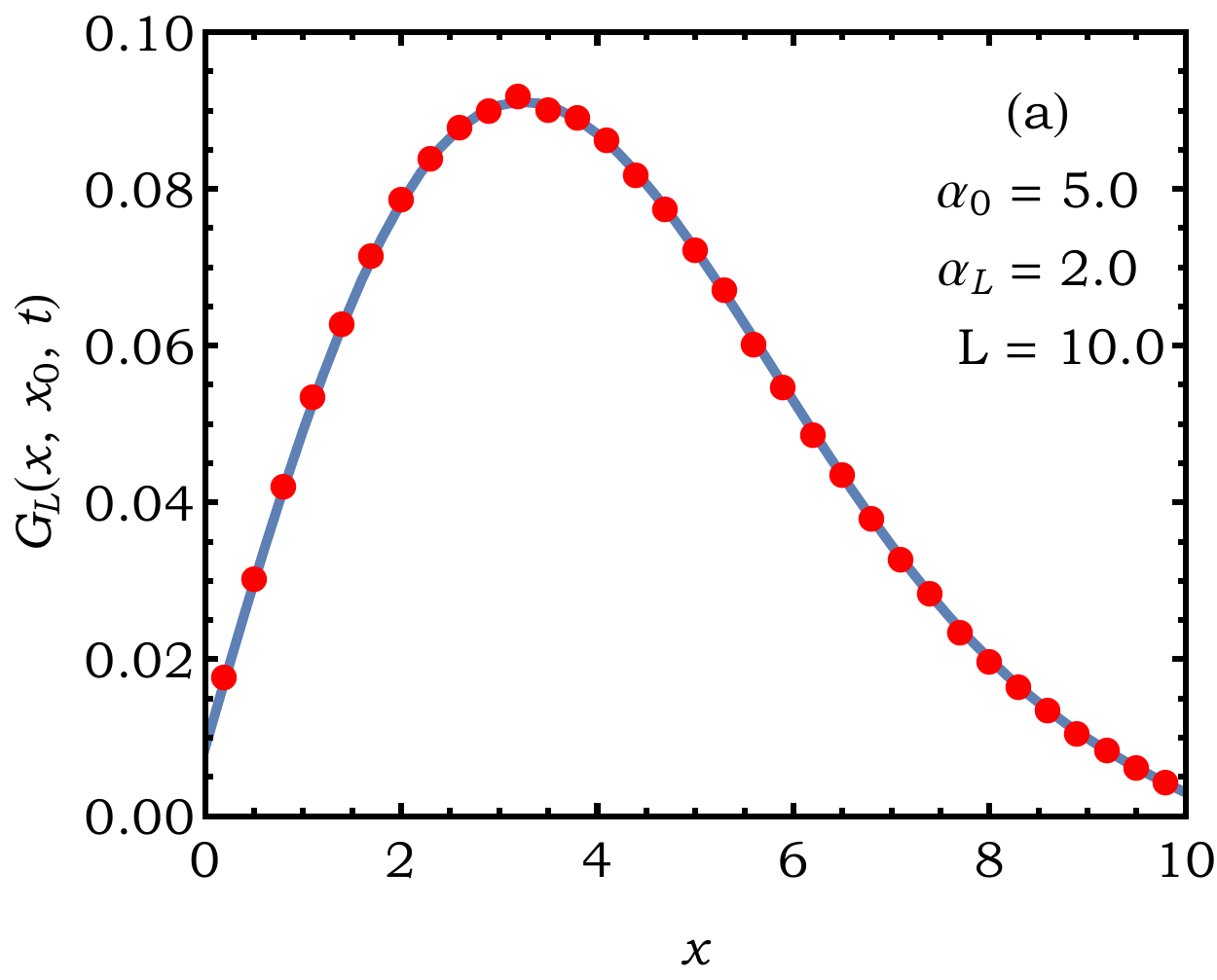}
\includegraphics[width=.394\hsize]{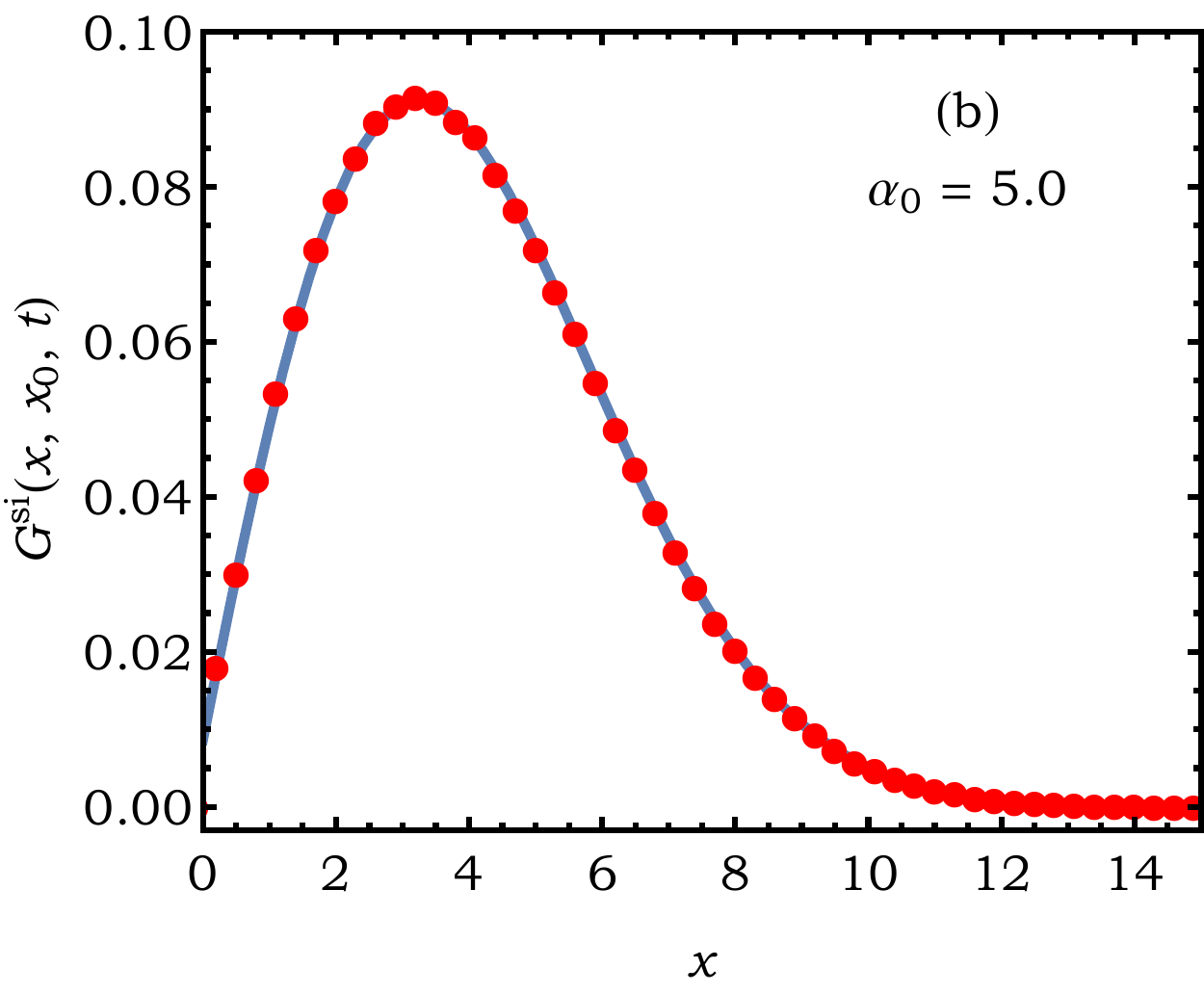}
\caption{ We have simulated the motion of a  Brownian molecule inside the interval  $[0,L]$ with reactive boundaries at $x=0$ and $x=L$. We have measured the displacement of the molecule
till an observation time $t=5$ for two individual cases. In the left panel (a), we show the probability distribution of the molecule displacement when it diffuses
strictly inside the box. We plot this numerical measurement (in red circles) with \eref{propagator-reactive-BC-full} (in solid line) and find an excellent match. 
In the right panel (b), we have considered the semi-infinite domain by taking the boundary at $L$ to be at infinity and measured the probability distribution of 
its displacement. The simulation data (in red circle) is plotted against the theoretical formula (in solid line) obtained using \eref{propagator-reactive-BC-semi-infinite}.
}
\l{propagators}
\end{figure}
\begin{figure}[h]
\includegraphics[width=.45\hsize]{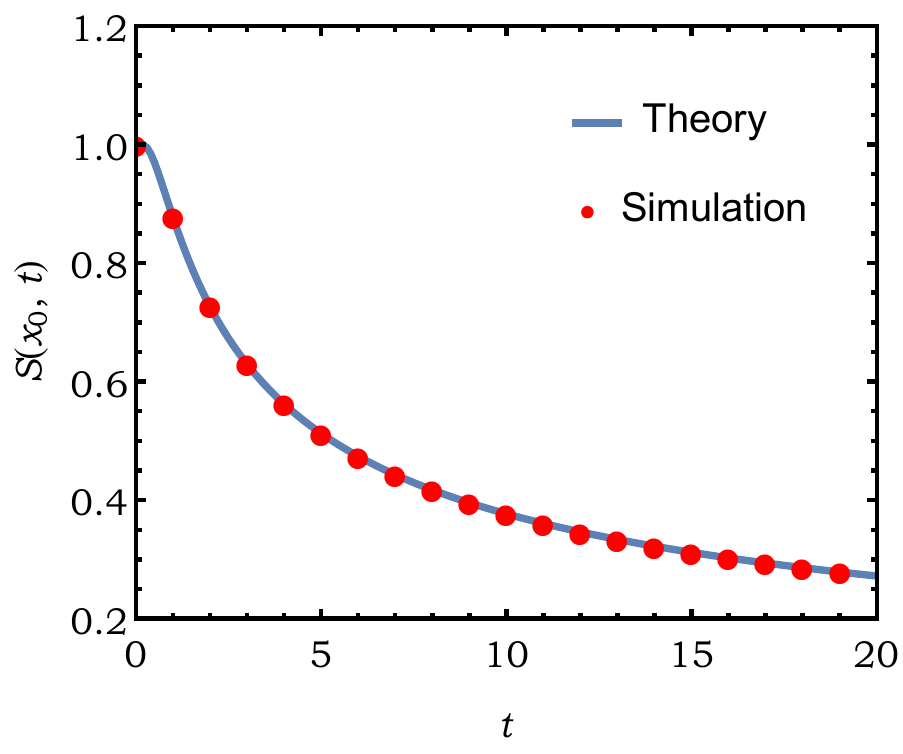}
\caption{Comparison of the expression \eref{survival-si} of the survival on the semi-infinite line with a 
reactive boundary at the origin with  an estimate  obtained from numerical simulations. We have used the following parameters for the simulation:
$\alpha_0=5.0$ and $x_0=2.0$. The simulation data (in red circles) is in excellent agreement with the analytical result (in solid line).}
\label{figsurv}
\end{figure}

\noindent
 Once we have  obtained an exact expression for the propagator $G_L(x,x_0,t)$ we can move on to derive the expressions for the survival probability and the mean absorption time.

{\it Survival probability}: The survival probability $S_L(x_0,t)$ inside the interval $[0,L]$ with reactive boundaries is a quantitative measure of the probability that the Brownian molecule
survives without being absorbed by  neither boundary. This probability  is obtained by integrating the final position $x$ from $0$ to $L$ in the expression of 
the propagator $G_L(x,x_0,t)$, yielding:
\bea
S_L(x_0,t)=\underset{k\in\mathcal{R}_k}{\sum}~e^{-Dk^2t}~\frac{\Big[\alpha_0(1-\cos(kL))+k \sin(kL)\Big]~[k \cos(kx_0)+\alpha_0 \sin(kx_0)]}{k(\alpha_0^2+k^2)\left[L+\f{\alpha_0}{\alpha_0^2+k^2}-\f{\alpha_L}{\alpha_L^2+k^2}\right]}\,.
\l{survival}
\eea
In the limit of $L \to \infty$ (semi-infinite domain), the survival probability at the boundary $0$ till an observation  time $t$ 
can  either be obtained from \eref{survival} , or
 by integrating out the final position $x$ in $G^{\text{si}}(x,x_0,t)$ from zero to infinity. Either way, the final result is
\bea
S(x_0,t)=e^{D \alpha_0^2 t+\alpha_0 x_0} ~\erfc \Big[ \f{x_0+2D\alpha_0 t}{\sqrt{4Dt}} \Big]+\erf \Big[\f{x_0}{\sqrt{4Dt}} \Big]\,,
\l{survival-si}
\eea
which, as shown in \fref{figsurv},  agrees  with estimates obtained  by Monte Carlo simulations.
It is easy to verify from \eref{survival-si} that when the boundary is completely reflective (i.e. $\alpha_0 \to 0$) the process always survives, that is 
$S(x_0,t)=1$. On the other hand, for a completely absorbing boundary condition (i.e. $\alpha_0 \to \infty$)
we readily recover the canonical result $S(x_0,t)=\erf \Big[x_0/\sqrt{4Dt} \Big]$ \cite{Redner}.  Next, we focus on the mean absorption time.

{\it Mean absorption time (MAT)}: 
The explicit form of the survival probability allows us to compute the first absorption time density immediately.
The distribution $f_L(t,x_0)$ of the absorption time is given by $f_L(t,x_0)=-dS_L(x_0,t)/dt$, from which the MAT  reads
\bea
T_L(x_0)=\underset{k\in\mathcal{R}_k}{\sum}~\frac{\Big[\alpha_0(1-\cos(kL))+k \sin(kL)\Big]~[k \cos(kx_0)+\alpha_0 \sin(kx_0)]}{Dk^3(\alpha_0^2+k^2)\left[L+\f{\alpha_0}{\alpha_0^2+k^2}-\f{\alpha_L}{\alpha_L^2+k^2}\right]}.
\l{FPT-full}
\eea
In the presence of two completely absorbing boundaries at $0$ and at $L$ (that is, by taking the limits $\alpha_0 \to \infty$ and $\alpha_L \to \infty$), the MAT  takes the following form
\bea
T_L^{\text{abs}}(x_0)=\frac{2L^2}{D \pi^3} \sum \limits_{n=1}^{\infty} \frac{1-(-1)^n}{n^3} \sin \left(\frac{n \pi x_0}{L}\right)=\frac{L^2}{2D }~z(1-z)~,
\eea
where $z=x_0/L$ and thus we recover this previously obtained result\cite{Redner, Satya-Krapivsky2010}. 
On the other hand, by taking the semi-infinite limit ($L \to \infty$) with finite $\alpha_0$ , the MAT diverges, as expected \cite{Redner}.

\begin{figure}[h]
\includegraphics[width=.44\hsize,valign=t]{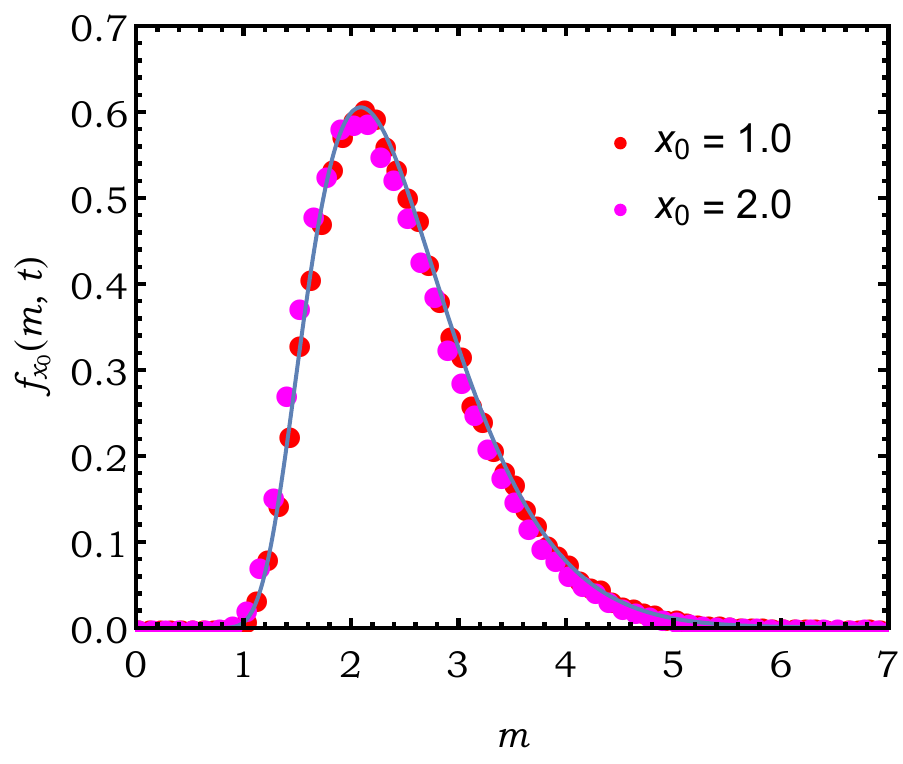}
\includegraphics[width=.44\hsize,valign=t]{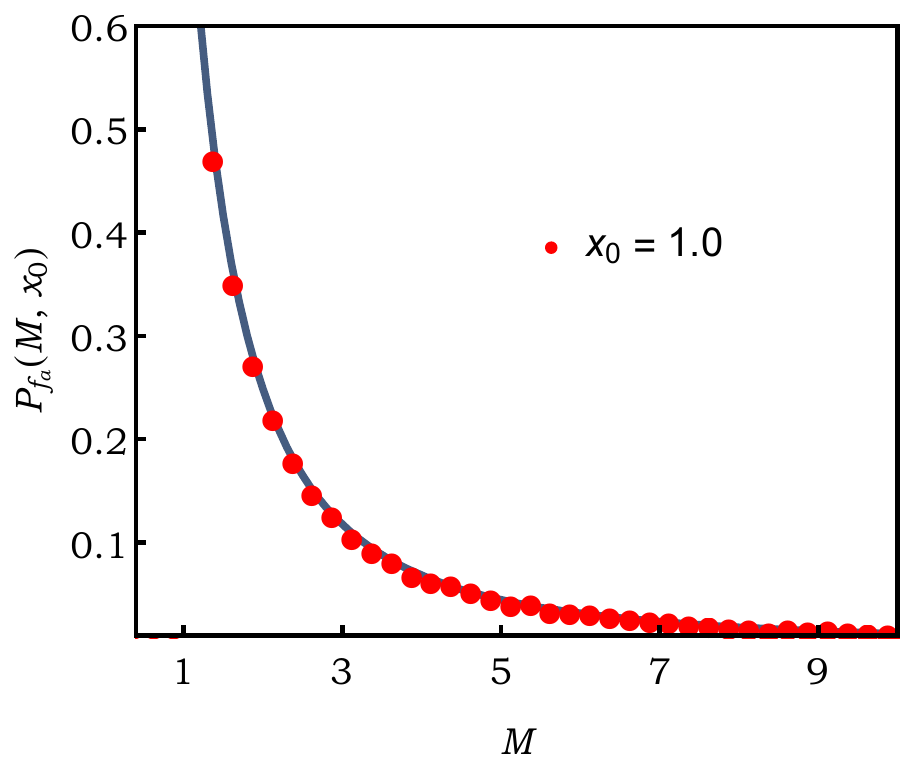}
\caption{{ Left panel: we have plotted  the scaled probability distribution $f_{x_0}(m,t)$ as a function of the scaled maximum displacement
$m$.
The point symbols represent data obtained from the Monte-Carlo simulations. The red and the magenta symbols indicate data
sets for $x_0=1.0$ and $x_0=2.0$, respectively. The solid line represents the theoretical curve from \eref{P-max-sclng}. 
An excellent data collapse for the scaled distribution is observed for different initial conditions, as predicted from our analysis.
{ Right panel}: we show the probability distribution of the maximum excursion $P_{f_a}(M,x_0)$  performed by the molecule before it is 
absorbed by the reactive boundary at the origin. 
The red circles are obtained from simulations and plotted against the solid curve obtained from the theoretical result using \eref{Pfa}. 
The parameters for this figure are: $D=1,~\alpha_0=5.0$, observation time $t=100.0$, and $x_0=1.0$.}
}
\label{fig1}
\end{figure}

\subsection{ Distribution of maximum displacement till a fixed time $t$}
\noindent
\l{max-displacement}
In this section, we study the statistics of the maximum displacement of our diffusing molecule. 
Often the amount of space visited by a diffusing chemical reagent inside  a cell becomes quite important as it might control the yield of a reaction.
One of the simplest measures of such excursions in one dimension is the maximum displacement traversed by the molecule.

In order to calculate the statistics of the maximum, it is convenient 
to  introduce the probability that the maximum displacement $M$ 
is less than or equal to $L$ (in 1D). To compute this  cumulative distribution,
one imposes an  adsorbing boundary 
at $x=L$ which is equivalent to taking $\alpha_L \to \infty$ in the original problem with two reactive boundaries.
Taking this limit in  \eref{propagator-reactive-BC-full}, we get the corresponding propagator
\begin{eqnarray}
\mathcal{G}_L(x,x_0,t) =\lim \limits_{\alpha_L\to \infty}G_L(x,x_0,t)
&=& \sum \limits_k 
\dfrac{[k \cos(kx)+\alpha_0 \sin(kx)]~[k \cos(kx_0)+\alpha_0 \sin(kx_0)]}{L(\alpha_0^2+k^2)+\alpha_0 } ~e^{-Dk^2t}, \nonumber \\
\text{with}  && ~~~~e^{2ikL}=\f{(k+i\alpha_0)}{(k-i\alpha_0)},
\label{green-one-mixed}
\end{eqnarray}
which was defined in \sref{Definitions}.
Integrating the final position $x$ of this propagator from $0$ to $L$, we get the associated survival probability $\mathcal{S}_L(x_0,t)$ of the particle:
\begin{equation}
\mathcal{S}_L(x_0,t) = \int_0^L dx~\mathcal{G}_L(x,x_0,t).
\end{equation}
Once we know $\mathcal{S}_L(x_0,t)$, the cumulative distribution of the maximum $M$ is obtained from the ratio (see \eref{maximum-cumulative})
\begin{equation}
H(L,t|x_0) = \text{Prob.}[M\le L,t |x_0] = \frac{\mathcal{S}_L(x_0,t)}{\mathcal{S}_{\infty}(x_0,t)}~~~
\text{where}~~~ \mathcal{S}_{\infty}(x_0,t) = \lim \limits_{L \to \infty} \mathcal{S}_L(x_0,t).
\label{cumulative-expression}
\end{equation}

\noindent
Clearly, in the limits $L\to \infty$ and $t \to \infty$, while keeping $\ell = \frac{L}{\sqrt{Dt}}$
 constant, the survival probability $\mathcal{S}_L(x_0,t)$ 
can be written in terms of the scaled variable $z_0 = \frac{x_0}{\sqrt{Dt}}$ such that $\mathcal{S}_L(x_0,t) = s_{\ell}(z_0,t)$.
After performing some algebraic manipulations one derives 
\begin{eqnarray}
 s_{\ell}(z_0,t) \simeq 2 \alpha_0 \sqrt{Dt}~\ell^2~\sum \limits_{n=1}^{\infty} 
 \frac{[1-(-1)^n]\left[ \frac{n\pi }{\ell} \cos \left(\frac{n\pi z_0}{\ell}\right)+\sqrt{Dt} \alpha_0\sin \left(\frac{n\pi z_0}{\ell}\right) \right]}
 {n\pi  \left[n^2\pi^2 + Dt~\alpha_0^2~\ell^2  + \sqrt{Dt}~\alpha_0~\ell\right] }~e^{-\frac{n^2\pi^2}{\ell^2}}.
\end{eqnarray}
To arrive at the above expression we  have replaced $kL=p\ell$ in \eref{green-one-mixed} to obtain the 
transcendental equation $\frac{p}{\tan(p\ell)} = - \alpha_0 \sqrt{Dt}$. For large $t$, the solution of this equation 
 is given approximately by  $p \approx \frac{n\pi}{\ell}$ with $n=1,2,3...$.
Hence by taking a derivative in \eref{cumulative-expression}, we obtain the following scaling form for the distribution $P_{x_0}(M,t)$ of the maximum  
\begin{equation}
 P_{x_0}(M,t) = \frac{1}{\sqrt{Dt}}~f_{x_0/\sqrt{Dt}} \left(\frac{M}{\sqrt{Dt}},t\right),~~~\text{where}~~
 f_{z_0}(m,t) = \frac{1}{\lim \limits_{\ell \to \infty}s_{\ell}(z_0,t)}\left(\frac{\partial s_{\ell}(z_0,t)}{\partial \ell} \right)_{\ell=m}~,
 \label{P-max-sclng}
\end{equation}
is the scaling function with the scaled maximum displacement $m=M/\sqrt{Dt}$. In the left panel of \fref{fig1}, we  have compared the expression of $\sqrt{Dt}P_{x_0}(M,t)$, given by \eref{P-max-sclng}, with  the 
corresponding estimate obtained by Monte Carlo simulations.
 As it can be appreciated  in \eref{P-max-sclng}, when plotted in this way we should observe a data collapse for various  values of the initial position $x_0$ 
and, indeed, this is what we have obtained numerically.

For an absorbing boundary at $x=0$ ( $\alpha_0 \to \infty$), the root equation in \eref{green-one-mixed} becomes $\tan(p\ell)=0$. 
This gives us the solutions $p \approx \frac{n\pi}{\ell}$ for $n=1,2,3...$, from which we derive the following form of the scaled survival probability 
\begin{eqnarray}
 s_{\ell}(z_0,t)=\frac{2}{\pi}\sum \limits_{n=1}^{\infty} \frac{1-(-1)^n}{n} ~\sin\left(\frac{n\pi z_0}{\ell}\right) ~e^{-\frac{n^2\pi^2}{\ell^2}}~,
\end{eqnarray}
which matches with the result derived previously in \cite{Satya-Krapivsky2010}.

\subsection{ Distribution of maximum displacement till the first absorption time}
\noindent
It  is also  fairly important to know  up to what extent the molecule has explored a region before it got absorbed at the reactive boundary. 
This can be quantified by the maximum displacement $M$ made by the molecule till the absorption time. 
The cumulative distribution $Q_{a}(L|x_0)=\text{Prob.}[M \leq L | x_0]$ of this maximum $M$ can be obtained from 
\begin{equation}
 Q_{a}(L|x_0) = D\int_0^{\infty}dt~\left( \frac{\partial \mathcal{G}_L}{\partial x}\right)_{x=0}.
\end{equation}
After  using the expression of $\mathcal{G}_L$ from \eref{green-one-mixed} and performing some manipulations we get 
\begin{eqnarray}
 Q_{a}(L|x_0) = \alpha_0 \sum_k \frac{[k \cos(kx_0)+\alpha_0 \sin(kx_0)]}{k[\alpha_0+L(\alpha_0^2+k^2)] },~~~~~\text{with}~~~
e^{2ikL}=\f{(k+i\alpha_0)}{(k-i\alpha_0)}.
\end{eqnarray}
Hence, the distribution of the maximum $P_{a}(M,x_0)$ till absorption time is obtained from 
\bea
 P_{f_a}(M,x_0) = \left(\frac{\partial  Q_{a}(L|x_0)}{\partial L} \right)_{L=M} \label{M-dist-a}.
\l{Pfa}
\eea
We have plotted \eref{Pfa} against  numerical simulations in \fref{fig1} (right panel) and we observe an excellent  agreement between them. 
In this case, one can also check the various limits which have been studied earlier. For example, in case of  a fully absorbing
boundary at $x=0$ ($\alpha_0 \to \infty$), one can show that the scaled cumulative distribution is given by 
\begin{eqnarray}
 Q_{a}(z_0,\ell)=2 \sum \limits_{n=1}^{\infty} \frac{\sin\left(\frac{n\pi x_0}{L}\right)}{n\pi}=1-\frac{x_0}{L}\,,
\end{eqnarray}
 which reproduces the result derived in  \cite{Satya-Krapivsky2010}.

\subsection{Reaction time spent at $y_0$ till observation time $t$:}
\l{reaction-time-profile}
\noindent
 Let us now focus our attention  at studying  the time spent by the molecule at some particular point while being observed for a time interval $t$. 
In the introduction we have mentioned examples where  the characterization of the reaction time renders subtle information 
about chemical or metabolic reactions. Here we show how one  can shed some light into  the statistical properties
of the reaction time for this paradigmatic model. For convenience, let us recall the definition of the reaction time
spent by a Brownian molecule at reaction coordinate $y_0$ given that it started  at $x_0$, or, in other words, 
\bea
L_t(y_0,x_0)=\int_0^t~dt'~\delta[ x(t')-y_0 |x_0]~,
\eea
where the observation time $t$ is fixed. Sometimes, this time can be taken stochastic, as we  discussed in \sref{reaction-time-till-ab}.

In order to compute the statistical properties of $L_t$, we introduce the generating function
\bea
\mathcal{Q}_p(y_0,x_0,t) = \langle e^{-p L_t(y_0,x_0)}\rangle  
\l{GF-LT}
\eea
where the average is performed with respect to the probability density $P(L_t|x_0,y_0,t)$
of the reaction time.
This average can be written explicitly after taking into account the exact path measure
\bea
\mathcal{Q}_p(y_0,x_0,t)=\f{1}{S_L(x_0,t)}~\int_0^L~dx~\int_{x(0)=x_0}^{x(t)=x} ~\mathcal{D}[x(\tau)]~e^{-  \int_0^t~d\tau \left[ 
 \frac{1}{4D}\left(\frac{dx(\tau)}{d\tau}\right)^2 + p\delta(x(\tau)-y_0)\right]} \,,
 \l{GF-LT-path-integral}
\eea
where the survival probability $S_L(x_0,t)$ weighs the surviving paths  (see \eref{survival}). 
Following  the Feynman-Kac method and  after introducing  an appropriate Hamiltonian $H_p$, one can map the original problem of evaluating the above path 
integral  into the computation of an imaginary time quantum propagator:
\bea
&\mathcal{Q}_p(y_0,x_0,t) &= \f{Q_p(y_0,x_0,t)}{S_L(x_0,t)}\,, 
\eea
where
\bea
 &Q_p(y_0,x_0,t) &= \int_0^{L}~ dx~ \langle x|e^{-t\hat{H}_p}|x_0 \rangle\,,
\l{reaction-time}
\eea
$\hat{H}_p(y_0) = -D\frac{d^2}{dx^2} + p~\delta(x-y_0)$, and $S_L(x_0,t)=\int_0^{L} ~dx~ \langle x|e^{-t\hat{H}_0}|x_0 \rangle$.
Using the backward Kolmogorov approach one can show that  $Q_p(y_0,x_0,t)$ obeys the following Fokker-Planck equation:
 \begin{eqnarray}
  \frac{\partial Q_p}{\partial t} = D\frac{\partial^2 Q_p}{\partial x_0^2} - p\delta(x_0-y_0)Q_p\,,
  \label{BFP-Q}
  \end{eqnarray}
 with  $Q_p(y_0,x_0,0) =1$, and boundary conditions
  \begin{eqnarray}
   ~~\left[\frac{\partial Q_p}{\partial x_0} - \alpha_0 Q_p \right]_{x_0=0}=0\,, \quad\quad
  \left[\frac{\partial Q_p}{\partial x_0} - \alpha_L Q_p \right]_{x_0=L}=0 \label{BFP-Q-BC}\,.
 \end{eqnarray}
The usual trick is then to 
write \eref{BFP-Q} in Laplace space with $\tilde{Q}_p(y_0,x_0,s)=\int_{0}^{\infty}dt~e^{-st}Q_p(y_0,x_0,t) $ such that
\bea
D \f{d^2 \tilde{Q}_p}{dx_0^2}-[s+ p ~\delta(x_0-y_0)]\tilde{Q}_p=-1 \l{BFP-Q_s}\,.
\eea 
The boundary conditions in
\eref{BFP-Q-BC} are automatically translated into
\bea
\left[\frac{\partial \tilde{Q}_p}{\partial x_0} - \alpha_0 \tilde{Q}_p \right]_{x_0=0}=0 ,\quad\quad
  \left[\frac{\partial \tilde{Q}_p}{\partial x_0} - \alpha_L \tilde{Q}_p \right]_{x_0=L}=0 \label{BFP-Q_s-BC}\,.
\eea
Although a formal solution of \eref{BFP-Q}  does exist,  the
exact Laplace inversion of $\tilde{Q}_p(y_0,x_0,s)$ turns out to be  rather difficult. 
However, we notice that 
a few exact results are available for the generating function $\mathcal{Q}_p(y_0,x_0,t)$ and for 
$P(L_t|x_0,y_0,t)$ when one approaches the limit of the semi-infinite domain. In the rest of this section we  thus focus 
on this particular limit. The interpretation of the generating function $\mathcal{Q}_p(y_0,x_0,t)$ 
is now extended accordingly to
\bea
\mathcal{Q}_p(y_0,x_0,t) &=& \f{Q_p(y_0,x_0,t)}{S(x_0,t)} \l{g-function}\,,\\
Q_p(y_0,x_0,t) &=& \int_0^{\infty}~ dx~ \langle x|e^{-t\hat{H}_p}|x_0 \rangle \l{Q_p-si}\,,
\eea
where $S(x_0,t)$ is given by \eref{survival-si}, while the governing equation for $\tilde{Q}_p(y_0,x_0,s)$ is still given by \eref{BFP-Q} 
with new boundary conditions 
\bea
\left[\frac{\partial \tilde{Q}_p}{\partial x_0} - \alpha_0 \tilde{Q}_p \right]_{x_0=0}=0 &,&\quad\quad
  \tilde{Q}_p(y_0,x_0,s)|_{x_0\to \infty}=\f{1}{s} ~.
  \l{BFP-Q_s-BC-si}
\eea
The second boundary condition is obtained from the fact that as $x_0\to\infty$, the local time $L_t\to 0$
which results  into $Q_p \to 1$. 
To obtain the complete solution of $\tilde{Q}_p(y_0,x_0,s)$, we need to solve \eref{BFP-Q} with the new boundary conditions  given by \eref{BFP-Q_s-BC-si}
in two different regions  separately. These are: (I) $0\le x_0 \le y_0$, and (II) $x_0 \ge y_0$.
A general solution in  these two regions can be written as 
\begin{eqnarray}
 \tilde{Q}_p^{(I)}(y_0,x_0,s)  &=& A ~e^{-x_0\sqrt{s/D}} + B ~e^{-x_0\sqrt{s/D}} + \frac{1}{s},~~~\text{for}~~0\le x_0 \le y_0 \label{sol-I-1} \\
 \tilde{Q}_p^{(II)}(y_0,x_0,s) &=& C ~e^{-x_0\sqrt{s/D}} + \frac{1}{s},~~~~~~~~~~~~~~~~~~~~~~~~~\text{for}~~x_0 \ge y_0.  \label{sol-I-2}
\end{eqnarray}
The constants $A$, $B$ and $C$ can be obtained from the following conditions 
\begin{itemize}
 \item Condition (a) : Boundary condition at $x_0=0$, i.e., $\frac{\partial \tilde{Q}_p}{\partial x_0}-\alpha_0 \tilde{Q}_p \big{|}_{x_0=0}=0$, which implies 
 \begin{equation}
  B(\sqrt{s/D}-\alpha_0) - A(\sqrt{s/D}+\alpha_0)=\frac{\alpha_0}{s}. \label{cond-a}
 \end{equation}

 \item Condition (b) : Continuity of the solution at $x_0=y_0$, which  implies
 \begin{equation}
 B e^{y_0\sqrt{s/D}} + A e^{-y_0\sqrt{s/D}} = C  e^{-y_0\sqrt{s/D}}. \label{cond-b}
 \end{equation}
Solving Eqs. (\ref{cond-a}) and (\ref{cond-b}) for $A$ and $B$, we get the following expressions in terms of $C$
\begin{eqnarray}
A e^{-y_0\sqrt{s/D}} &=& \frac{\sqrt{D}}{2R_{y_0}(s)} \left(-\frac{\alpha}{s} + C~(\sqrt{s/D}-\alpha_0)~e^{-2y_0\sqrt{s/D}} \right), \label{A} \\ 
B e^{y_0\sqrt{s/D}} &=& \frac{\sqrt{D}}{2R_{y_0}(s)} \left(\frac{\alpha}{s} + C~(\sqrt{s/D}+\alpha_0) \right),~~~\text{where}, \label{B} \\
 R_{y_0}(s) &=&\sqrt{s} \cosh(y_0\sqrt{s/D}) + \alpha_0 \sqrt{D} \sinh(y_0\sqrt{s/D}). \label{R_x_0}
\end{eqnarray}

 \item Condition (c) : Discontinuity of the first derivative at  $x_0=y_0$, i.e., 
 \begin{equation}
  \frac{\partial  \tilde{Q}_p^{(II)}}{\partial x_0}\big{|}_{x_0=y_0} -  \frac{\partial  \tilde{Q}_p^{(I)}}{\partial x_0}\big{|}_{x_0=y_0} = \frac{p}{D} \tilde{Q}_p^{(II)}(y_0,y_0,s). \label{dis-continuity}
 \end{equation}
Using the expressions of $A$ and $B$ from Eqs. (\ref{A}) and (\ref{B}) in the above equation we get 
\begin{equation}
 C ~e^{-y_0\sqrt{s/D}} = -\frac{1}{s}~\frac{pR_{y_0}(s) + \alpha_0D\sqrt{s}}{pR_{y_0}(s) + \sqrt{s}(\sqrt{sD}+\alpha_0D)~e^{y_0\sqrt{s/D}}}. \label{C}
\end{equation}

\end{itemize}
Using this expression of $C$ in Eqs. (\ref{A}) and (\ref{B}) we get explicit expressions of $A$ and $B$. As a result we have  a complete specification of 
$\tilde{Q}_p(y_0,x_0,s)$, from which performing double inverse Laplace transformation one can, in principle, obtain $P(L_t|x_0,y_0,t)$ for any reaction location $y_0$. In the following 
we consider two choices for this location to demonstrate few exact results. The choices are: (i) at the origin ($y_0=0$),  and (ii) at its initial position  $y_0=x_0$. 

\subsubsection{Reaction time around the reactive boundary $y_0=0$}
\label{x_0-0}
\noindent
In this case, $R_{y_0}(s)=\sqrt{s}$. Hence, the constant $C$ in \eref{C} now reads 
\begin{equation}
 C= -\frac{1}{s}~\frac{p + \alpha_0D}{p +\alpha_0D + \sqrt{sD}},
\end{equation}
 and as a result the function $\tilde{Q}_p(0,x_0,s)$ has the  following form 
\begin{eqnarray}
 \tilde{Q}_p(0,x_0,s) &=& \frac{1}{s} -\frac{1}{s}~\frac{p + \alpha_0D}{p +\alpha_0D + \sqrt{sD}}~e^{-x_0\sqrt{s/D}}, \nonumber \\
 &=& \frac{1-e^{-x_0\sqrt{s/D}}}{s} + \frac{e^{-x_0\sqrt{s/D}}}{\sqrt{s}(p +\alpha_0D + \sqrt{sD})}\,.
\end{eqnarray}
Performing the inverse Laplace transform with respect to $s$ we obtain
\begin{eqnarray}
Q_p(0,x_0,t) &=& S(x_0,t)~\mathcal{Q}_p(0,x_0,t)
 = e^{D\alpha_p^2t+\alpha_px_0}\text{erfc}\left(\frac{2D\alpha_pt+x_0}{\sqrt{4Dt}} \right) + \text{erf}\left(\frac{x_0}{\sqrt{4Dt}} \right),
\end{eqnarray}
where $\alpha_p=\alpha_0+\frac{p}{D}$. Now performing the  inverse
Laplace transform with respect to $p$, we get 
\begin{equation}
q_0(L_t,t|x_0)=\mathcal{L}^{-1}_{L_t}(Q_p(0,x_0,t)) = 
2~\text{erf}\left(\frac{x_0}{\sqrt{4Dt}} \right)~\delta(L_t) + \sqrt{\frac{D}{\pi t}}~e^{-\alpha_0 L_t D} ~e^{-\frac{(x_0+L_t D)^2}{4Dt}}~. \label{LT-y_0-0}
\end{equation}
Hence, the distribution of the local time (density) at $y_0=0$ is given by  
\begin{eqnarray}
 P(L_t|x_0,0,t)&=& \frac{1}{S(x_0,t)}
 ~\left[ 2~\text{erf}\left(\frac{x_0}{\sqrt{4Dt}} \right)~\delta(L_t) + \sqrt{\frac{D}{\pi t}}~e^{-\alpha_0 L_t D} ~e^{-\frac{(x_0+L_t D)^2}{4Dt}}\right], \nonumber \\
 \text{where} && S(x_0,t)=e^{D\alpha_0^2t+\alpha_0x_0}\text{erfc}\left(\frac{2D\alpha_0t+x_0}{\sqrt{4Dt}} \right) + \text{erf}\left(\frac{x_0}{\sqrt{4Dt}} \right).
 \label{S}
\end{eqnarray}
The $\delta(L_t)$ term in  \eref{LT-y_0-0} arises from those paths which   are absorbed at the reactive boundary upon their first passage.
The factor $\text{erf}({x_0}/{\sqrt{4Dt}})/S(x_0,t)$ represents the fraction of those paths, which starting  at $x_0$ survived
being absorbed by  the reactive boundary and, moreover, did not make any visit to the boundary till time $t$. Note 
that this is the term that survives in the $\alpha_0 \to \infty$ limit, i.e., when the boundary becomes completely absorbing.

\begin{figure}[h]
\includegraphics[scale=0.9]{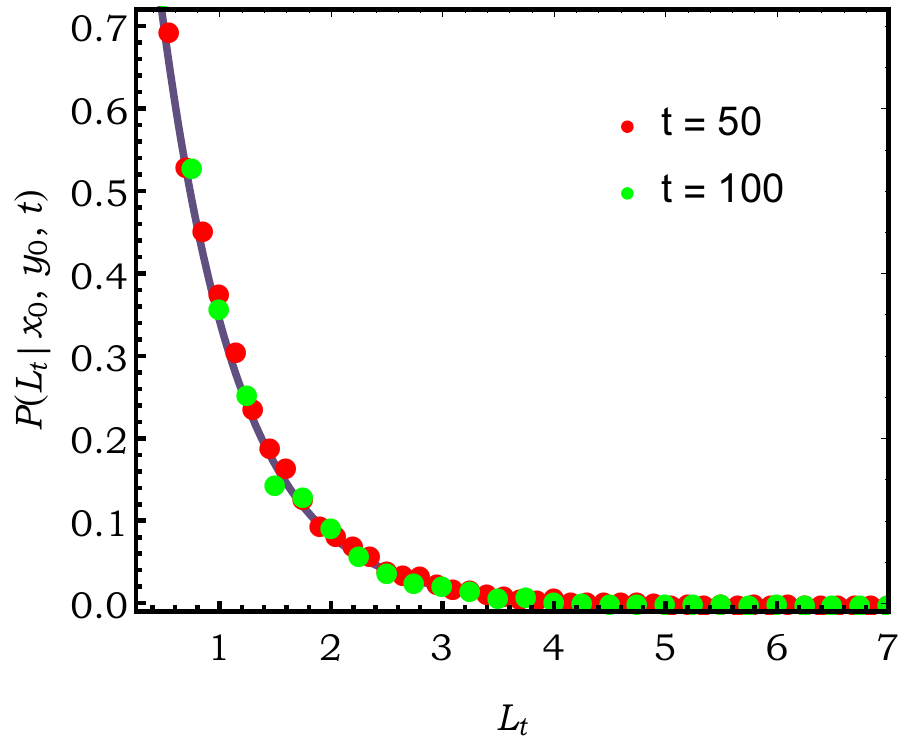}
\caption{Numerical distribution of reaction time $L_t$ till the observation time $t$, starting from $x_0$. We measure the reaction time around its initial
coordinate i.e., $y_0=x_0$ here.
The point symbols (red and green) represent simulation data obtained 
for time $(t=50$ and $100)$ respectively. The solid line is obtained from the theoretical result in \eref{pdf-ell-lrg-t}. Since \eref{pdf-ell-lrg-t} does not depend on $t$,
we see all the data fall in the same line. The parameters for this figure are: $D=1, y_0=x_0=0.5$, and $\alpha_0=5.0$.}
\label{fig6}
\end{figure}

\subsubsection{Reaction time around its initial position $y_0=x_0$}
 Putting $y_0=x_0$ in Eqs. \eqref{sol-I-1} and  \eqref{sol-I-2}, and using the expression of $C$ from \eref{C} we have 
\begin{equation}
  \tilde{Q}_p(x_0,x_0,s) = \frac{1}{s} + C~e^{-x_0\sqrt{s/D}} = \frac{1}{s}~\frac{s\sqrt{D}+\alpha_0D\sqrt{s}(1-e^{-y_0\sqrt{s/D}})}{R_{y_0}(s)e^{-x_0\sqrt{s/D}}}
  ~\times~\frac{1}{p + \frac{s\sqrt{D}+\alpha_0D\sqrt{s}}{R_{x_0}(s)}e^{-x_0\sqrt{s/D}} },
\end{equation}
where $R_{x_0}(s)$ is given by \eref{R_x_0}. Performing the inverse Laplace transform with respect to $p$ we get 
\begin{equation}
\tilde{Z}_s(L_t,x_0) = \mathcal{L}^{-1}_{L_t}(\tilde{Q}_p(x_0,x_0,s)) 
= \frac{1}{s}~\frac{s\sqrt{D}+\alpha_0D\sqrt{s}(1-e^{-x_0\sqrt{s/D}})}{R_{x_0}(s)e^{-x_0\sqrt{s/D}}}~\times~
e^{-\frac{s\sqrt{D}+\alpha_0D\sqrt{s}}{R_{x_0}(s)}e^{-x_0\sqrt{s/D}}~L_t}. \label{Z_s}
\end{equation}
We need to perform another inverse Laplace  transform with respect to $s$ to get $Z_t(L_t,x_0)=\mathcal{L}^{-1}_s(\tilde{Z}_s(L_t,x_0))$. 
We can  then use this  result to obtain the reaction time density $P(L_t|x_0,x_0,t)$ as 
\begin{equation}
 P(L_t|x_0,x_0,t) = \frac{Z_t(L_t,x_0)}{S(x_0,t)},
\end{equation}
where $S(x_0,t)$ is the survival probability given  by \eref{S}. 
It seems  to  be rather cumbersome to perform the second Laplace inversion analytically. However, it is possible to look at the following limits :
\begin{itemize}
 \item \underline{Large $t$ :}  In this case, the dominant contribution to the inverse Laplace transform with respect to $s$ comes from the small $s$ limit of 
 $\tilde{Z}_s(L_t,x_0)$ and that is given by 
 \begin{equation}
  \tilde{Z}_s(L_t,x_0)|_{s \to 0} \simeq \sqrt{D}~e^{-\frac{\alpha_0D L_t}{1+x_0\alpha_0}}~\frac{1}{\sqrt{s}}~e^{-\frac{D L_t \sqrt{s}}{1+x_0\alpha_0}},
 \end{equation}
which provides 
\begin{equation}
Z_t(L_t,x_0)|_{t\to \infty} \simeq \sqrt{\frac{D}{\pi t}}~e^{-\frac{\alpha_0D L_t}{1+x_0\alpha_0}}~e^{-\frac{D^2 L_t^2}{4t(1+x_0\alpha_0)^2}}.
\end{equation}
On the other hand $S(x_0,t)|_{t\to \infty} \simeq \frac{1+x_0\alpha_0}{\sqrt{\pi Dt\alpha_0^2}}$. Hence the density function $P(L_t|x_0,x_0,t)$ for large $t$ is 
given by 
\begin{eqnarray}
 P(L_t|x_0,x_0,t)|_{t\to \infty} &\simeq& \frac{D\alpha_0}{1+x_0\alpha_0}
 ~e^{-\frac{\alpha_0D }{1+x_0\alpha_0}L_t}~e^{-\frac{D^2 }{4t(1+x_0\alpha_0)^2}L_t^2}, \nonumber \\ 
 &\simeq&  \frac{D\alpha_0}{1+x_0\alpha_0} ~e^{-\frac{\alpha_0D }{1+x_0\alpha_0}L_t}. \label{pdf-ell-lrg-t}
\end{eqnarray}
Note that the distribution of the local time becomes independent of time for asymptotically large $t$. In \fref{fig6}, we  have compared the analytical 
expression of $ P(L_t|x_0,x_0,t)$   given by \eref{pdf-ell-lrg-t}
 to the density obtained from the direct simulation of the Langevin equation. We see  a nice agreement
between them and, moreover, we also observe that the 
densities are independent of $t$, as predicted from \eref{pdf-ell-lrg-t}.


\item \underline{Small $t$ :} In this case, the dominant contribution to the inverse Laplace transform with respect to $s$ comes from the large $s$ limit of 
 $\tilde{Z}_s(L_t,x_0)$ and that is given by
  \begin{equation}
  \tilde{Z}_s(L_t,x_0)|_{s \to \infty} \simeq \sqrt{D}~\frac{e^{-L_t \sqrt{s D}}}{\sqrt{s}},
 \end{equation}
 which provides 
\begin{equation}
Z_t(L_t,x_0)|_{t\to 0} \simeq \sqrt{\frac{D}{\pi t}}~e^{-\frac{L_t^2D}{4t}}. 
\end{equation}
On the other hand $S(x_0,t)|_{t\to 0} \simeq 1- \frac{4\alpha_0 (Dt)^{3/2}}{\sqrt{\pi}~x_0^2}~e^{-\frac{x_0^2}{4Dt}}$. 
Therefore, the density function $P(L_t|x_0,x_0,t)$ for small $t$  reads 
\begin{equation}
 P(L_t|x_0,x_0,t)|_{t\to 0} \simeq \sqrt{\frac{D}{\pi t}}
 ~\frac{e^{-\frac{L_t^2D}{4t}}}{ 1- \frac{4\alpha_0 (Dt)^{3/2}}{\sqrt{\pi}~x_0^2}~e^{-\frac{x_0^2}{4Dt}}}~ 
 \simeq~ \sqrt{\frac{D}{\pi t}}~e^{-\frac{L_t^2D}{4t}}. \label{pdf-ell-smll-t}
\end{equation} 
Note that the above expression is independent of the reactive constant $\alpha_0$, since, in the small time limit, the system is yet to see the boundaries
and would behave like a free diffusion \cite{Sanjib-Satya}.
\end{itemize}

\subsection{Reaction time spent at $y_0$ till the absorption time:}
\label{reaction-time-till-ab}
\noindent
In the preceding section, we have studied the reaction time profile of a molecule inside the cell for a fixed duration $t$.
 It may, however, occur that
the molecule is absorbed before the reaction takes place (with zero contribution to the reaction time profile) or the reaction occurs 
with an immediate adsorption and there is no need to perform the experiment for the whole duration $t$.
This motivates us to study the reaction time profile till the adsorption event. 
Since, the absorption time is a functional of the trajectory, the reaction time
profile is accounted by two stochastic terms:  the noise and the random absorption time.

\begin{figure}[t]
\includegraphics[scale=.4]{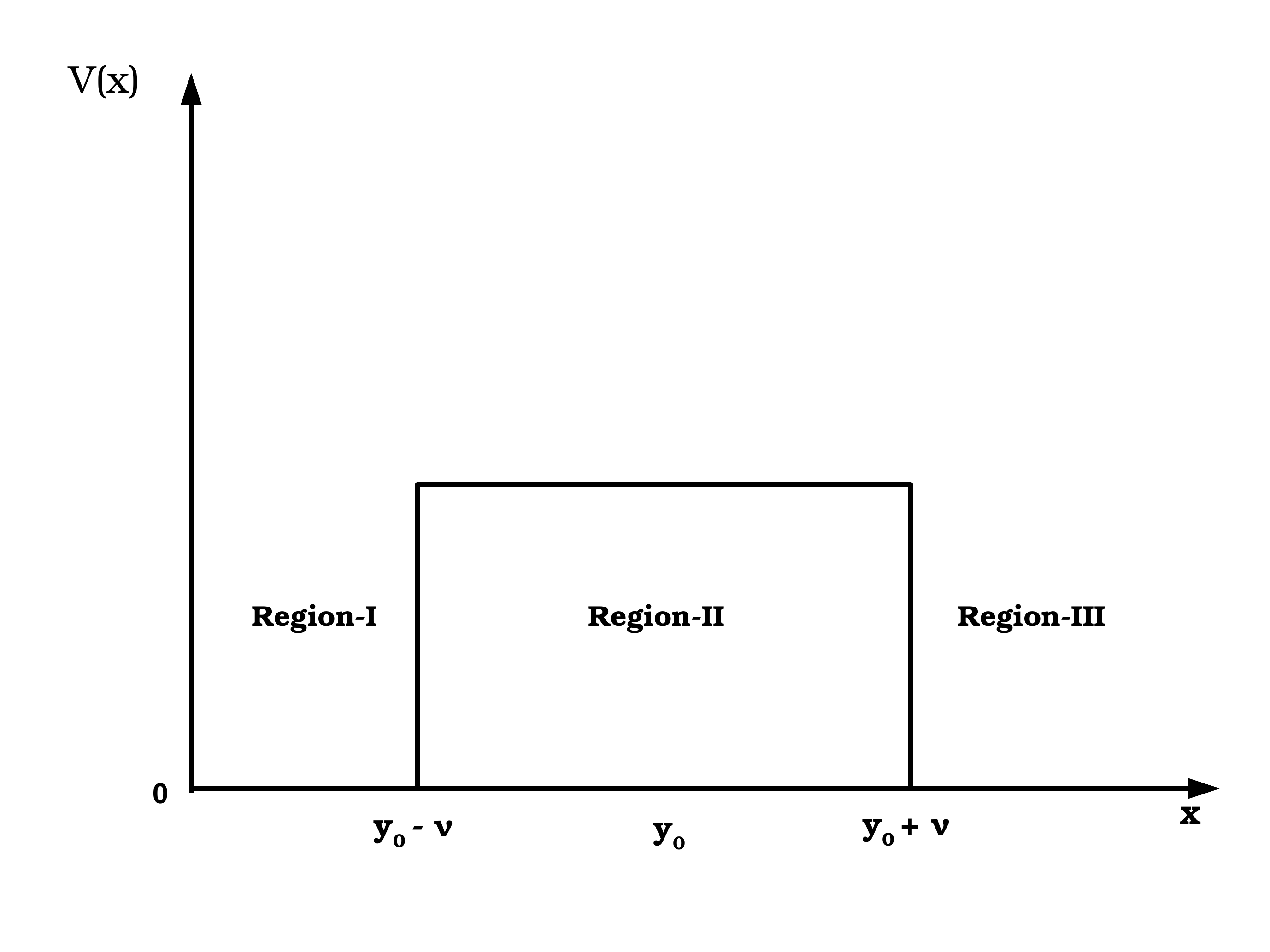}
\caption{We have shown the construction of the effective potential $V(x)$ as in \eref{potential} and the disjoint regions as \eref{potential} indicates.}
\label{fig4}
\end{figure}

Let us define the local time (density) $L_a(y_0,x_0)$ till the adsorption event as
\begin{equation}
 L_a(y_0,x_0) = \int_0^{t_a} \delta[x(t)-y_0|x_0]dt,
 \label{localT-till-ab-def}
\end{equation}
where $t_a$ is the time when the molecule is absorbed at the boundary $0$ and the initial condition is set as
$x(0)=x_0$. Clearly, the time $t_a$ is a stochastic quantity
and this kind of functional is often known as the first passage time functionals in the literature \cite{Majumdar:2005}.
It will prove convenient to rewrite 
the reaction time $L_a(y_0,x_0)$ in the following way
\bea
 L_a(y_0,x_0) = \lim \limits_{\nu \to 0} ~\frac{W_{\nu}(y_0,x_0)}{2\nu}~,
 \l{eff-pot}
\eea
where
\bea
W_{\nu}(y_0,x_0) = \int_0^{t_a}~V[x(t)] ~dt,~~
 \text{with}~~V(x)=\Theta[y_0+\nu-x]~\Theta[x-y_0+\nu]~.
 \label{potential}
\eea
The function $\Theta[x]$ represents the Heaviside  step function. 
In \fref{fig4}, we give a sketch of the
effective potential $V(x)$.
In fact,
$W_{\nu}(y_0,x_0)$ represents the time spent by the particle inside the box, centered around $y_0$, till the absorption time $t_a$. Hence taking
$\nu \to 0$ justifies our construction in \eref{eff-pot} along with \eref{potential}.

As done in the previous section,  we start  once again with the generating function $Q_{\nu}(p,y_0,x_0)$ of $W_{\nu}(y_0,x_0)$ which is defined as 
\begin{equation}
 Q_{\nu}(p,y_0,x_0) = \langle e^{-pW_{\nu}(y_0,x_0)}\rangle=\langle e^{-p\int_0^{t_a}~V(x(t)) dt} \rangle~.
\end{equation}
The generating function $\mathcal{T}_p(y_0,x_0)$ associated to $L_a$ is equivalently defined as 
\begin{equation}
 \mathcal{T}_p(y_0,x_0) = \langle e^{-pL_a(y_0,x_0)}\rangle_{x_0}=\langle e^{-p\int_0^{t_a}~\delta[x(t) - y_0|x_0] dt}\rangle~.
 \l{reaction-time-Laplace}
\end{equation}
Both generating functions are  related to each other by
\begin{equation}
 \mathcal{T}_p(y_0,x_0) = \lim \limits_{\nu \to 0} Q_{\nu}(p/2\nu,y_0,x_0). \label{mcalT}
\end{equation}
Using the Markov property one can show \cite{Majumdar:2005} that, $Q_{\nu}(p,y_0,x_0)$ 
satisfies the following differential equation 
\begin{equation}
D\frac{d^2 Q_{\nu}}{d x_0^2} - p~V(x_0) ~Q_{\nu}=0, \label{BFP-Qa} 
\end{equation}
 which is accompanied by the following boundary conditions
: as $x_0 \to \infty$, the time $t_a$ to get absorbed  also tends to  infinity, which implies that 
$Q_{\nu}(p,y_0,x_0)$  cannot diverge and, secondly,  at  $x=0$, 
we have a reactive boundary, which implies that
\begin{equation}
 \left[\frac{d Q_{\nu} }{d x_0} - \alpha_0~Q_{\nu}(p,y_0,x_0) \right]_{x_0=0} = -\alpha_0. \label{BC-y0-0}
\end{equation}
The solution of the  differential equation  \eref{BFP-Qa}  is naturally divided 
in three different regions: (I) $0 \le x_0 \le y_0-\nu$, (II) $y_0-\nu \le x_0 \le y_0+\nu$, and 
(III) $x_0 \ge y_0+\nu$  (see \fref{fig4}). The solution can be written in the following way
\begin{eqnarray}
 Q_{\nu}(p,y_0,x_0) &=& A_{\nu} + B_{\nu}~x_0,~~~~~~~~~~~~~~~~~~~~~~~~~~~~~~~~~~\text{for}~~~0 \le x_0 \le y_0-\nu, \label{sol_I} \\
 Q_{\nu}(p,y_0,x_0) &=& F_{\nu}~\cosh \left[(x_0-y_0-\nu)\sqrt{\frac{p}{D}} ~\right],~~~~\text{for}~~~y_0-\nu \le x_0 \le y_0+\nu \label{sol_II} \\
  Q_{\nu}(p,y_0,x_0) &=& F_{\nu},~~~~~~~~~~~~~~~~~~~~~~~~~~~~~~~~~~~~~~~~~~~~~~~\text{for}~~~ x_0 \ge y_0+\nu. \label{sol_III}
\end{eqnarray}
The constants $A_{\nu}$, $B_{\nu}$ and $F_{\nu}$ are computed from the following matching conditions: (a) continuity of the solutions, 
(b) continuity of their derivatives at $x_0=y_0\pm \nu$ and 
(c) using the reactive boundary condition at $x_0=0$ according to \eref{BC-y0-0}. Now,
\begin{itemize}
 \item Using  the matching conditions (a) and (b) imply
 \begin{eqnarray}
  A_{\nu} + B_{\nu}(y_0-\nu)-F_{\nu}~\cosh \left[2\nu \sqrt{\frac{p}{D}}~\right] &=& 0, \label{cond-1} \\
   B_{\nu}(y_0-\nu)\sqrt{D}+\sqrt{p}F_{\nu}~\sinh \left[2\nu \sqrt{\frac{p}{D}}~\right] &=& 0, \label{cond-2}
 \end{eqnarray}

 \item and using  the boundary condition (c) at $x_0=0$ we finally obtain 
 \begin{equation}
  B_{\nu}=\alpha_0~A_{\nu} - \alpha_0. \label{cond-3}
 \end{equation}

\end{itemize}
Solving the three equations (\ref{cond-1}), (\ref{cond-2}) and (\ref{cond-3}), we get 
\begin{eqnarray}
 A_{\nu}(p) &=& \alpha_0~\frac{(y_0-\nu)\sqrt{p}\sinh (2\nu \sqrt{p/D}) + \sqrt{D}\cosh (2\nu \sqrt{p/D})}
 {[1+\alpha_0(y_0-\nu)]\sqrt{p}\sinh (2\nu \sqrt{p/D}) + \alpha_0\sqrt{D}\cosh (2\nu \sqrt{p/D})}, \label{A_nu} \\
 B_{\nu}(p) &=&-\alpha_0~\frac{\sqrt{p}\sinh (2\nu \sqrt{p/D}) }
 {[1+\alpha_0(y_0-\nu)]\sqrt{p}\sinh (2\nu \sqrt{p/D}) + \alpha_0\sqrt{D}\cosh (2\nu \sqrt{p/D})}, \label{B_nu} \\
 F_{\nu}(p) &=& \frac{\alpha_0\sqrt{D} }
 {[1+\alpha_0(y_0-\nu)]\sqrt{p}\sinh (2\nu \sqrt{p/D}) + \alpha_0\sqrt{D}\cosh (2\nu \sqrt{p/D})}. \label{F_nu}
\end{eqnarray}

\begin{figure}[t]
\centering
\includegraphics[width=.4\hsize,valign=t]{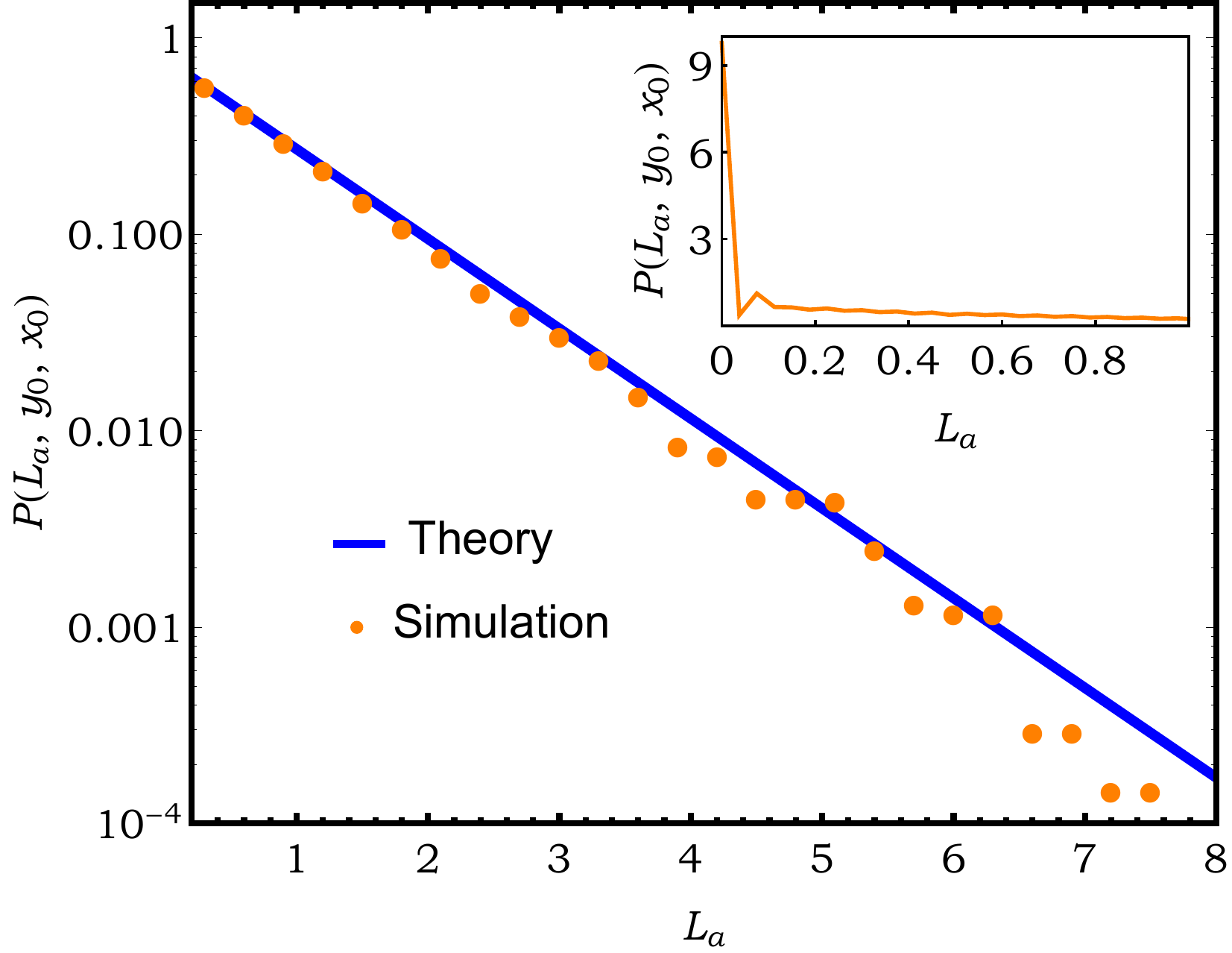}
\includegraphics[width=.415\hsize,valign=t]{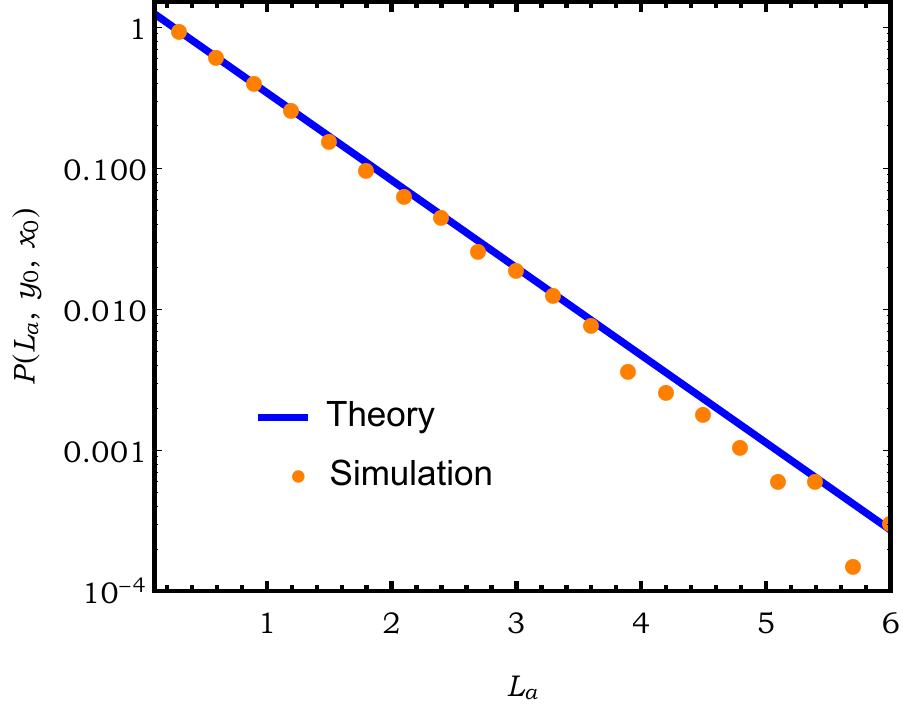}
\caption{Numerical distribution of the reaction time $P(L_a,y_0,x_0)$ at $y_0$ till its first adsorption at $0$ given that
it had started from $x_0$. We have compared the numerical simulation with the distribution obtained analytically in \eref{P-ell-y0-lt-x0}. The distribution has different analytical
forms depending on $x_0, y_0$. In the left panel, we have considered $x_0=0.5, y_0=0.75$ (so that $x_0 < y_0$) while in the right panel we have 
$x_0=0.75, y_0=0.5$ (so that $x_0 > y_0$). In both plots, analytical formulas (solid blue lines) have been plotted against the simulation curves (marked with orange circles).
The inset in the left panel shows the presence of the Dirac delta function when $x_0<y_0$.
 The parameters for this figure are set as: $D=1, \alpha_0=5.0$. }
\label{fig5}
\end{figure}

\noindent
To establish the connection between the generating function as prescribed in \eref{mcalT}, we first
take  the $\nu \to 0$ limit  in the expressions of the constants $A_{\nu}(p/2\nu)$, $B_{\nu}(p/2\nu)$ and $F_{\nu}(p/2\nu)$. This yields 
\begin{eqnarray}
 A_a &=& \lim \limits_{\nu \to 0}A_{\nu}(p/2\nu) = \frac{\alpha_0(y_0p+D)}{(1+\alpha_0y_0)p+\alpha_0D}, \label{A_a} \\
 B_a &=& \lim \limits_{\nu \to 0}B_{\nu}(p/2\nu) = -\frac{\alpha_0p}{(1+\alpha_0y_0)p+\alpha_0D}, \label{B_a} \\
 F_a &=& \lim \limits_{\nu \to 0}F_{\nu}(p/2\nu) = \frac{\alpha_0D}{(1+\alpha_0y_0)p+\alpha_0D}. \label{F_a}
\end{eqnarray}
 All in all, we arrive to the  following solution of the generating function $\mathcal{T}_p(y_0,x_0)$ 
 \bea
 \mathcal{T}_p(y_0,x_0)= 
\begin{cases}
    \frac{p~\alpha_0(y_0-x_0)+\alpha_0 D}{(1+\alpha_0y_0)p+\alpha_0D},& \text{if~~} 0 \le x_0 \le y_0, \label{mcalT_I}\\
    \frac{\alpha_0D}{(1+\alpha_0y_0)p+\alpha_0D},              & \text{if~~}  x_0 \ge y_0. \label{mcalT_II}
\end{cases}
\eea
 Finally, to obtain the distribution $P(L_a,y_0,x_0)$ we need to perform the inverse Laplace transform in \eref{mcalT_I}
 , which yields the full distribution of the reaction time profile 
\bea
 P(L_a,y_0,x_0) =
 \begin{cases}
  \frac{2\alpha_0(y_0-x_0)}{1+\alpha_0y_0}~\delta(L_a) 
 + \alpha_0D~\frac{1+\alpha_0 x_0}{(1+\alpha_0y_0)^2}~e^{-\frac{\alpha_0D}{1+\alpha_0y_0}~L_a},& \text{if~~}0\le x_0 \le y_0, \label{P-ell-y0-lt-x0} \\
 \frac{\alpha_0D}{(1+\alpha_0y_0)}~e^{-\frac{\alpha_0D}{1+\alpha_0y_0}~L_a}, & \text{if~~}  x_0 \ge y_0.\label{P-ell-y0-gt-x0}
 \end{cases}
\eea
 In the above expression we see  that there is a $\delta$-function contribution to the distribution
of $L_a$. This  results from those paths which  are 
absorbed at the boundary before making a first passage to $y_0$.  For this reason, the $\delta$-function
contribution appears only when $0 \le x_0 \le y_0$  or, in other words, 
there is a  chance for the molecule to  be absorbed at the boundary at $x=0$ 
before ever reaching $y_0$ for the first time. This element of chance decreases as one starts 
closer to $y_0$, which explains the multiplicative factor $(y_0-x_0)$. One the other hand,
if the molecule starts at $x_0  \ge y_0$ it will  definitely cross $y_0$ before it  is 
absorbed at $x=0$ and, as a result, it spends some time around $y_0$ (due  to the Brownian nature 
of the motion).  This is the reason behind the absence of a $\delta$-function term for $x_0  \ge y_0$. 
In \fref{fig5} we compare the analytical expression  given by \eref{P-ell-y0-gt-x0} 
  to simulation results 
finding, once again, an excellent agreement.

\section{Methods of simulations}
\l{simulations}
In this section, we outline the method we have used to simulate our system. There are many ways to generate trajectories of a
Brownian particle diffusing in a box with two reactive boundaries \cite{Singer08,Erban07,Andrews-2009,Bray-2004}. 
In one such method, as described in \cite{Singer08}, the authors
define the partially reflected process as the limit of a Markovian jump process generated by the dynamics using  an Euler scheme.
Using boundary layer analysis, they derive a relation between the reactive constants and the reflection probability. In another paper \cite{Erban07},
the authors study four different approaches to simulate such systems and they have derived the correct choices of the reactive boundary conditions to implement in stochastic simulations.
In this paper, we have adapted one of these approaches from \cite{Erban07} to generate the trajectories. This is known as the Euler scheme for velocity jump process \cite{Erban07}. 
In this scheme, one simulates the system by defining an auxiliary underdamped motion (by introducing a velocity component along with the existing position component)
with friction coefficient $\Gamma$.
The dynamics is discrete in time, continuous in space and discontinuous in velocities. We sketch the basic steps in the following  lines.

Let us consider a system of $N$ independent molecules (i.e., N independent and identical copies of the system). The $i$-th molecule is described by two variables: its position $x_i(t)$ and  velocity $v_i(t)$
at time $t$. The underdamped dynamics for the set $\{ x_i(t),v_i(t) \}$ at each time step $\Delta t$ is introduced in the following way:
\bea
x_i(t+\Delta t) &=& x_i(t)+v_i(t)\Delta t, \nn \\
v_i(t+\Delta t) &=& v_i(t)- \Gamma v_i(t)\Delta t+ \Gamma~ \sqrt{2D \Delta t} ~\eta_i,
\l{Sim1}
\eea
where $\Gamma$ should be taken large and $\eta_i$ is a normally distributed random variable with zero mean and unit variance \cite{Erban07}. 
In this problem we have two reactive boundaries at $x=0$ and $x=L$ and
the reactive boundary conditions can be stated as follows: whenever a molecule hits any one of the two boundaries ($0,L$)
it is adsorbed with probability $p_0/\sqrt{\Gamma}$ or $p_L/\sqrt{\Gamma}$ respectively,
 or reflected otherwise. The implementation is the following: whenever the value $x_i(t+\Delta t)$ computed from \eref{Sim1} is negative then
\bea
x_i(t+\Delta t) &=& -x_i(t)-v_i(t)\Delta t, \nn \\
v_i(t+\Delta t) &=& -v_i(t)+ \Gamma v_i(t)\Delta t- \Gamma~ \sqrt{2D \Delta t} ~\eta_i,
\eea
with probability $1-\f{p_0}{\sqrt{\Gamma}}$, or we remove the $i$-th molecule from the system. 
On the other hand, if $x_i(t+\Delta t)$ computed from \eref{Sim1} is greater than $L$, we do the following
\bea
x_i(t+\Delta t) &=& 2 L-x_i(t)-v_i(t)\Delta t, \nn \\
v_i(t+\Delta t) &=& -v_i(t)+ \Gamma v_i(t)\Delta t- \Gamma~ \sqrt{2D \Delta t} ~\eta_i,
\eea
with probability $1-\f{p_L}{\sqrt{\Gamma}}$, otherwise we remove the $i$-th molecule from the system. Finally, we use the following relation between the 
reactive constants and the reflection probabilities \cite{Erban07}
\bea
p_0 &=& \f{\alpha_0 \sqrt{2\pi}}{\sqrt{D}} \nn \\
p_L &=& \f{\alpha_L \sqrt{2\pi}}{\sqrt{D}}.
\eea
It is important that only in the high friction limit, that is, only when $\Gamma$ is large enough,
we recover the diffusion equation  \eref{propagator-reactive-BC} which is the overdamped limit and the inception of our study. The above prescription allows
us to successfully generate Brownian trajectories in the presence of two reactive boundaries and
the number of molecules present in the system after a given time $t$ is simply proportional to the probability density defined in \eref{propagator-reactive-BC}.
We conclude this section by stating that
 other statistical quantities such as  the survival probability  or 
 the reaction time profile can  be also simulated using  this method.

\section{Conclusions and Future Outlook}
\l{conclusion} 
In this paper, we have built a comprehensive theory to study various statistical properties of a Brownian molecule in presence of reactive boundaries. 
Such boundaries are ubiquitous
in physics, chemistry and biology. Several molecular movements inside a cell can 
 fairly well be  described by a Brownian motion where the cell boundary provides the confined geometry. One often considers these
boundaries to be either completely reflecting or completely absorbing. However, the effects of adsorption, catalysis etc. occurring at the cell
boundary makes them reactive, in the sense that these boundaries are neither fully absorbing nor fully reflecting. In this paper, we have looked 
at the Brownian motion of a molecule in one dimension with partially absorbing (reflecting) boundaries. In this case, we find that the propagator
of the molecule is different from that with fully absorbing/reflecting boundaries. We have also looked at the survival properties of the molecule, 
which also provides explicit expressions of the distribution of absorption time,  the mean absorption time as well as the distribution of the 
maximum displacement. Using the Feynman-Kac formalism, we have investigated the distribution of the reaction or local time density both when observed for a
fixed time or till the absorption time. We have obtained explicit expressions of the distribution of the reaction time which give an excellent match with 
the numerical simulations.

Our work can be extended in multiple directions. 
In a recent study \cite{Grebenkov-17-target}, the authors  considered the mean first passage time to a reaction event on a specific site in a cylindrical 
geometry with mixed boundary conditions. It would be interesting to estimate the survival probabilities and the longest excursions (maximum displacement
and time to reach the maximum) in such set up and further extend it to different non-uniform geometries.
It would be also interesting to see how the properties of
a tagged particle in presence of other particles  are effected  by considering
reactive crossing conditions. 
Effects of a partially absorbing boundary have also been investigated recently in an interesting stochastic dynamics namely stochastic resetting \cite{EM11a,EM11b,KMSS14,Pal14}
which mixes long range moves along with the local moves due to diffusion \cite{Whitehouse-Martin-Satya}. Moreover, such dynamics could be
quite benifical strategies to target search \cite{SR2016,AAM16,PalShlomi2016}. It is left for future studies to combine this dynamics in conjugation with diffusion
to expedite first passage processes to a target in a confined domain with reactive boundaries.

\section{Acknowledgment}
Arnab Pal gratefully acknowledges support from the Raymond and Beverly Sackler Post-Doctoral Scholarship at Tel-Aviv University.
Anupam Kundu acknowledges support from DST grant under project No. ECR/2017/000634. This work benefited from the support of the
project 5604-2 of the Indo-French Centre for the Promotion of Advanced Research (IFCPAR). We would also like to thank the Weizmann Institute of Science
for the hospitality during the SRitp workshop where part of this work was done. Isaac P\'erez Castillo thanks hospitality of the Laboratoire de Physique Th\'eorique et Mod\`eles Statistiques (Universit\'e de Paris-Sud), where this work was initiated.

\section{Appendix}
\l{appendix}

\subsection{Derivations of propagator in semi-infinite domain}
\l{propagator in semi-infinite domain}
\noindent
To derive the propagator in the semi-infinite domain, we have to take the limit $L \to \infty$
in \eref{propagator-reactive-BC-full}. In this limit \eref{k-equation} yields 
$k=\f{n\pi}{L}$. The summation over $k$ can now be converted into an integral
over $k$ as $L \to \infty$. A short calculation gives us
\begin{align}
G_t^{\text{si}}(x,x_0)&=\f{1}{2\pi}~\left[\f{\partial}{\partial x}\f{\partial}{\partial x_0}+\alpha_0\left( \f{\partial}{\partial x}+\f{\partial}{\partial x_0}\right)+\alpha_0^2\right]
\int_{-\infty}^{\infty}~dk~\f{2 \sin(kx_0) \sin(kx)}{\alpha_0^2+k^2}~e^{-Dk^2t} \nonumber \\
&=\f{1}{2\pi}~\Big(\f{\partial}{\partial x}+\alpha_0\Big)\Big(\f{\partial}{\partial x_0}+\alpha_0\Big)
\int_{-\infty}^{\infty}~dk~\f{\cos[k(x+x_0)]+\cos[k(x-x_0)]}{\alpha_0^2+k^2}~e^{-Dk^2t} \,.
\l{G-semi-infinite}
\end{align}
To compute \eref{G-semi-infinite} we consider the following integral
\bea
I_{\alpha}(z)&=&\int_{-\infty}^{\infty}~dk~\f{ \cos(kz)}{\alpha^2+k^2}~e^{-k^2t} \nn \\
&=&e^{\alpha^2t} \left[ \int_{-\infty}^{\infty}~dk~ \f{ \cos(k z)}{\alpha^2+k^2}~e^{-(\alpha^2+k^2)t} \right]\,.
\l{integral-I}
\eea
By noting that $\int_{t}^{\infty}~dt'~e^{-(k^2+\alpha^2)t'}=\f{e^{-(\alpha^2+k^2)t}}{\alpha^2+k^2}$, we 
find from \eref{integral-I}
\bea
I_{\alpha}(z)&=& e^{\alpha^2t}~\int_{t}^{\infty}~dt'~e^{-\alpha^2t'}~\text{Re}\left[ \int_{-\infty}^{\infty}~dk~e^{ik z}~e^{-k^2t'} \right] \nn \\
&=& \f{2\sqrt{\pi}}{\alpha}~e^{y^2}~\int_{y}^{\infty}~dq~e^{-q^2}~e^{-\f{z^2 \alpha^2}{4q^2}}\,,
\l{integral-II}
\eea
where $y=\alpha \sqrt{t}$. Applying the following standard integral formula
\bea
\int_{y}^{\infty}~dt~e^{-a^2t^2-\f{b^2}{t^2}}=\f{\sqrt{\pi}}{4a}~e^{2ab}~\erfc \left[ ay+\f{b}{y} \right]
+\f{\sqrt{\pi}}{4a}~e^{-2ab}~\erfc \left[ ay-\f{b}{y} \right]\,,
\eea
in \eref{integral-II}, eventually we obtain from \eref{integral-I}
\bea
I_{\alpha}(z)=\f{\pi}{2\alpha}e^{\alpha^2 t}~\left[ e^{z\alpha}~\erfc \left( \f{2\alpha t+z}{\sqrt{4t}} \right) +
e^{-x_0\alpha}~\erfc \left( \f{2\alpha t-z}{\sqrt{4t}} \right)\right]\,.
\eea
Plugging the above expression in \eref{G-semi-infinite}, we obtain \eref{propagator-reactive-BC-semi-infinite} 
as mentioned in the main text along with \eref{psi-definition}.


\begin{thebibliography}{1}

\bibitem{Berg83} H. Berg,  Random Walks in Biology (Princeton: Princeton University Press) (1983).

\bibitem{Crank75} J. Crank, The Mathematics of Diffusion (Oxford: Oxford University Press) (1975).

\bibitem{Cooper} G. M. Cooper and R. E. Hausman, The Cell: A Molecular Approach, (Seventh Edition: Sinauer Associates) (2016).

\bibitem{Vroman} L. Vroman, A. L. Adams, G. C. Fischer, and P. C. Munoz, Blood \textbf{55}, 156 (1980); J. L. Brash et al., ibid. \textbf{71}, 932 (1988); 
J. G. Donaldson, R. A. Kahn, J. Lippincott-schwartz, and R. D. Klausner, Science \textbf{254}, 1197 (1991).

\bibitem{Bychuk} O. V. Bychuk and B. O. Shaughnessy, Phys. Rev. Lett. {\bf 74}, 1795 (1995); R. Valiullin, R. Kimmich, and N. Fatkullin, 
Phys. Rev. E \textbf{56}, 4371 (1997).

\bibitem{Grebenkov06} D. S. Grebenkov, in Focus on Probability Theory, Ed. L. R. Velle, pp. 135-169 (Nova Science Publishers) (2006).

\bibitem{GrebenkovRMP07} D. S. Grebenkov, Rev. Mod. Phys. \textbf{79}, 1077 (2007).

\bibitem{Zwanzig90} R. Zwanzig,  Proc. Nati. Acad. Sci. USA, \textbf{87},  5856-5857, (1990).

\bibitem{Allison85} S. A. Allison, S. H. Northrup, and J. A. McCammon, The Journal of Chemical Physics \textbf{83}, 2894 (1985).

\bibitem{Lamm83} G. Lamm and K. Schulten, The Journal of Chemical Physics \textbf{78}, 2713 (1983).

\bibitem{Chapman16}  S. J. Chapman, R. Erban, and S. A. Isaacson, Siam J. Appl. Math., \textbf{76}(1), 368-390 (2016).

\bibitem{Erban07}  R. Erban and  S. J. Chapman, Phys. Biol. \textbf{4},  16-28 (2007).

\bibitem{Hattne05} J. Hattne, D. Fagne and J. Elf,  Bioinformatics {\bf 21}, 2923-4 (2005).

\bibitem{Naqvi82} K. Naqvi, K. Mork, and S. Waldenstrom, Phys. Rev. Lett. {\bf 49}, 304-7 (1982).

\bibitem{Jacob-2012} E. Jacob, Stochastic Processes and their Applications, {\bf 122}(1), pp.191-216 (2012).

\bibitem{El-Sh-2000} M. A El-Shehawey, Journal of Physics A: Mathematical and General {\bf 33},  49, 9005 (2000).

\bibitem{Goodrich-54} F. C. Goodrich, Journal of Chemical Physics, {\bf 22} 588-594, (1954).

\bibitem{Grebenkov-review} D. S. Grebenkov, Journal of Physics A: Mathematical and Theoretical {\bf 48}, 1, 013001 (2014).


\bibitem{Vrentas89} J. S.Vrentas, C.M. Vrentas , Chem. Eng. Sci., \textbf{44}(12), 3001 - 3003 (1989).

\bibitem{Isaac}I. P. Castillo and T. Dupic, J. Stat. Phys. \textbf{156}, 606  (2014).

\bibitem{Hippel} P. H. von Hippel and O. G. Berg, J. Biol. Chem. \textbf{264}, 675 (1989), and references therein.

\bibitem{Slutsky} M. Slutsky and L. A. Mirny, Biophys. J. {\bf 87}, 4021 (2004); M. Coppey, O. Benichou, R. Voituriez, and M. Moreau, Biophys. J. {\bf 87}, 1640 (2004); 
I. M. Sokolov, R. Metzler, K. Pant, and M. C. Williams, Biophys. J. \textbf{89}, 895 (2005); Y. M. Wang, R. H. Austin, and E. C. Cox, Phys. Rev. Lett. \textbf{97}, 048302 (2006).

\bibitem{Grebenkov15} D. S. Grebenkov, Phys. Rev. E \textbf{91}, 052108 (2015).

\bibitem{Dickman01} R. Dickman and D. ben-Avraham, Phys. Rev. E \textbf{64}, 020102(R) (2001).

\bibitem{Skorokhod59} A. V. Skorokhod, Siam Theory Probab. Appl., {\bf 6} (3), 264-274. (1959); ibid. {\bf 7} (1), 3-23 (1959).

\bibitem{Burdzy04} K. Burdzy, Z.Q. Chen, and J. Sylvester, Ann. Probab., {\bf 32}, 1B , 775-804 (2004).

\bibitem{Singer08} A. Singer, Z. Schuss, A. Osipov, and D. Holcman, SIAM J. Appl. Math., {\bf 68} (3), 844-868 (2008).

\bibitem{Batsilas03} L. Batsilas, A. M. Berezhkovskii, and S. Y. Shvartsman, Biophys. J., \textbf{85} ,  3659-3665 (2003).

\bibitem{Berezhkovskii} A. M. Berezhkovskii, Y. A. Makhnovskii, M. I. Monine, V. Yu. Zitserman, and S. Y. Shvartsman, J. Chem. Phys., \textbf{121},  11390-11394 (2004).

\bibitem{Monine} M. I. Monine and J. M. Haugh,  J. Chem. Phys., \textbf{123} , 074908 (2005).

\bibitem{Metzler07} M. A. Lomholt, I. M. Zaid, and R. Metzler, Phys. Rev. Lett. \textbf{98}, 200603 (2007).

\bibitem{Stapf06}  S. Stapf, R. Kimmich, and R. O. Seitter, Phys. Rev.Lett. \textbf{75}, 2855 (1995); P. Levitz et al., ibid. \textbf{96}, 180601 (2006).

\bibitem{Sapoval5} B. Sapoval, Phys. Rev. Lett. \textbf{73}, 3314 (1994).

\bibitem{Sonin} A. A. Sonin, A. Bonfillon, and D. Langevin, Phys. Rev. Lett. \textbf{71}, 2342 (1993); C. Stenvot and D. Langevin, Langmuir \textbf{4}, 1179 (1988).

\bibitem{Weibel} E. R. Weibel, The Pathway for oxygen. Structure and function in the mammalian 
respiratory system (Harvard University Press, Cambridge, Massachusetts and London, England) (1984).

\bibitem{Filoche}  B. Mauroy, M. Filoche, E. R. Weibel, and B. Sapoval, Nature \textbf{427}, 633 (2004).

\bibitem{Sapoval1} B. Sapoval, M. Filoche, and E. R. Weibel, Branched Structures, Acinus Morphology and Optimal Design of Mammalian Lungs, in Branching in
nature, Eds. by V. Fleury, J.-F. Gouyet, and M. Leonetti,  225-242 (EDP Sciences/Springer Verlag) (2001).


\bibitem{Sapoval2} B. Sapoval, J. S. Andrade Jr., and M. Filoche, Chem. Eng. Sci. \textbf{56}, 5011 (2001).

\bibitem{Sapoval3} J. S. Andrade Jr., M. Filoche, and B. Sapoval, Europhys. Lett. \textbf{55}, 573 (2001).

\bibitem{Sapoval4} J. S. Andrade Jr., H. F. da Silva, M. Baquil, and B. Sapoval, Phys. Rev. E \textbf{68}, 041608 (2003).

\bibitem{BMS13} A. J. Bray, S. N. Majumdar, and G. Schehr, Advances in Physics \textbf{62}, 225 (2013).

\bibitem{Majumdar:2005} S. N. Majumdar, Curr. Sci.  \textbf{89}, 2076 (2005).

\bibitem{Redner} S. Redner,  A guide to First-Passage Processes, (Cambridge University Press, Cambridge), (2001).

\bibitem{First-Passage-Book} R. Metzler, G. Oshanin, S. Redner Ed., First-Passage Phenomena and Their Applications, (World Scientific) (2014).

\bibitem{Benichou2011RMP} O. B\'{e}nichou, C Loverdo, M. Moreau, and R. Voituriez, Rev. Mod. Phys. 83 {\bf 81} 2011.


\bibitem{Grebenkov17} D. S. Grebenkov and J.-F. Rupprecht, The Journal of Chemical Physics \textbf{146}, 084106 (2017).

\bibitem{Salminen11} P. Salminen, and M. Yor, Periodica Mathematica Hungarica, {\bf 62}, 1, 75-101 (2011).

\bibitem{Ben-naim93}E. Ben-Naim, S. Redner and G. H. Weiss, J. Stat. Phys.  {\bf 71}, 75 (1993).

\bibitem{Whitehouse-Martin-Satya} J. Whitehouse, M. R. Evans, and S. N. Majumdar, Phys. Rev. E {\bf 87}, 022118 (2013).

\bibitem{Sano79}H. Sano and M. Tachiya, The Journal of Chemical Physics \textbf{71}, 1276 (1979).

\bibitem{Michaelis-Menten} L. Menten and M. I. Michaelis, Biochemische Zeitschrift \textbf{49}, 333 (1913).

\bibitem{Sanjib-Satya} S. Sabhapandit, S. N. Majumdar, and A. Comtet, Phys. Rev. E {\bf 73}, 051102 (2006).

\bibitem{Donsker-Varadhan} M. D. Donsker and S. R. S. Varadhan, I, Commun. Pure Appl. Math. \textbf{28}, 1 (1975).

\bibitem{Barato} A. C. Barato and R. Chetrite, J. Stat. Phys. \textbf{160}, 1154 (2015).
 

\bibitem{Satya-Krapivsky2010} P. L. Krapivsky, S. N. Majumdar, A. Rosso, J. Phys. A: Math. Theor. \textbf{43}, 315001  (2010).


\bibitem{Andrews-2009} S. S. Andrews, Physical Biology, {\bf 6}(4):046015, (2009).

\bibitem{Bray-2004} S. S. Andrews and D. Bray, Physical biology, {\bf 1} (3), p.137 (2004).



\bibitem{Grebenkov-17-target} D. S. Grebenkov, R Metzler, G Oshanin, New Journal of Physics {\bf 19} (10), 103025 (2017).

\bibitem{EM11a} M. R. Evans, and S. N. Majumdar, Phys. Rev. Lett. \textbf{106}, 160601 (2011).

\bibitem{EM11b} M. R. Evans, and S. N. Majumdar, J. Phys. A: Math. Theor. \textbf{44}, 435001 (2011).

\bibitem{KMSS14} L. Kusmierz, S. N. Majumdar, S. Sabhapandit, and G. Schehr, Phys. Rev. Lett. \textbf{113}, 220602 (2014).

\bibitem{Pal14} A. Pal,  Phys. Rev. E \textbf{91}, 012113 (2015).

\bibitem{SR2016} S. Reuveni, Phys. Rev. Lett. \textbf{116}, 170601 (2016).

\bibitem{AAM16} A. Pal, A. Kundu and M. R. Evans, J. Phys. A: Math. Theor. {\bf 49},  225001 (2016).

\bibitem{PalShlomi2016} A. Pal, and S. Reuveni, Phys. Rev. Lett. {\bf 118}, 030603 (2017).



\end{thebibliography}
\end{document}